\documentclass{tMOP2e}

\usepackage{graphicx}
\usepackage{booktabs}

\newcommand{\re}{\mathop{\mathrm{Re}}}

\newcommand{\D}{\mathop{\mathrm{d}}}
\newcommand{\I}{\mathop{\mathrm{i}}}

\begin{document}
\title{Coherence properties of the radiation from FLASH}

\author{E.A. Schneidmiller, M.V. Yurkov, DESY, Hamburg, Germany }

\maketitle

\begin{abstract}

FLASH is the first free electron laser user facility operating in the vacuum
ultraviolet and soft x-ray wavelength range. Many user experiments require
knowledge of the spatial and temporal coherence properties of the radiation. In
this paper we present an analysis of the coherence properties of the
radiation for the fundamental and for the higher odd frequency harmonics.
We show that temporal and spatial coherence reach maximum close to the FEL saturation but may degrade significantly
in the post-saturation regime. We also find that the pointing stability of short FEL pulses is limited due to the fact
that non-azimuthal FEL eigenmodes are not sufficiently suppressed.
We discuss possible ways for improving the degree of transverse coherence and the pointing stability.

\end{abstract}

\section{Introduction}

FLASH (Free electron LASer in Hamburg) operates in the vacuum-ultraviolet and
soft X-ray range between approximately 45 nm and 4.2 nm wavelength.
This facility originated from the TESLA
Test Facility (TTF) project which was built to test the technology for the
linear collider TESLA (TeV Energy Superconducting Linear Collider)
\cite{ttf-cdr,tesla-cdr,tesla-tdr}. At the same time the project of the TTF
free electron laser (TTF FEL) has been launched aiming a minimum radiation
wavelength of 6 nm \cite{ttf-fel-cdr}. The first stage of the project
successfully generated VUV light in the year 2000 \cite{ttf-fel-prl-2000}. High
power (a few GW) and ultrashort (a few 10 fs) radiation pulses have been used
in pioneer user experiments
\cite{ttf-sat-prl,ttf-sat-epj,mueller-nature,jacek-ablation}. It has been
decided later on to transform the TTF FEL to the dedicated FEL user facility
which is in operation since 2004 under the name of FLASH
\cite{vuvfel-exp,flash-13nm,flash-2012,flash-2013-1,flash-2013-2,flash-2014-1,flash-2014-2}.
FLASH free electron laser is driven by 1.25 GeV superconducting linear
accelerator. Five scientific  instruments have been in use since the
commissioning of the facility in 2004. Second stage, FLASH2 is under
commissioning now. First lasing at FLASH2 has been obtained in August, 2014
\cite{flash-2014-1,flash-2014-2}. FLASH facility is also used for the
development and testing of technology for the European XFEL and for the
International Linear Collider (ILC)
\cite{tesla-tdr-xfel,euro-xfel-tdr,ilc-tdr,weise-2014}.

With the present undulator (period 2.73 cm, peak field 0.486 T) the minimum
wavelength of 4.2 nm is determined by the maximum electron beam energy of
approximately 1.25 GeV. There is a tendency for users at FLASH to extend
wavelength range to shorter wavelengths. The first target is the
so-called water window, i.e. the range between the K-Absorption edges of carbon
and oxygen at 4.38nm and 2.34 nm, respectively. Currently a minimum wavelength of
FLASH is just below the carbon edge. Another range of interest refers to the
edges of magnetic elements which spans below water window. Higher odd harmonics
of SASE radiation can be used to generate radiation at such short wavelengths.
Pioneer experiment for studying magnetic materials using FEL radiation has been
performed at FLASH at 1.6 nm wavelength, the 5th harmonic of the fundamental at
8 nm \cite{gutt-2009}. Many user's experiments rely on coherent properties of
the radiation, both temporal and spatial. This relates not only to the
fundamental harmonic, but to the higher odd harmonics as well
\cite{vart1,vart2}.

FLASH is single pass free electron laser starting from the shot noise in the
electron beam \cite{ks-1980,dks-1982,pel-1985}. This FEL amplifier
configuration is frequently named as SASE FEL (Self Amplified Spontaneous
Emission FEL \cite{boni-sase}). Previous studies have shown that coherence
properties of the radiation from SASE FEL strongly evolve during
the amplification process
\cite{coherence-oc,coherence-njp,coherence-anal-oc,coh-fel2012,trcoh-oc}. At
the initial stage of amplification the spatial coherence is poor, and the
radiation consists of a large number of transverse modes
\cite{trcoh-oc,moore-modes,fs2r-1993,xie-modes,ssy-eigen-bet,book,kim-inval-1,wang-yu,yu-krinsky}.
Longitudinal coherence is poor as well
\cite{bonifacio-sase,bonifacio-sase-prl,stat-oc}. In the exponential stage of
amplification the transverse modes with higher gain dominate over modes with lower gain when
the undulator length progresses. This feature is also known as the mode
competition process. Longitudinal coherence is also improving in the high gain
linear regime \cite{stat-oc,kim,harm-prst}. Mode selection process stops at the
onset of the nonlinear regime, and maximum values of the degree of the
transverse coherence and of the coherence time are reached at this point.
Undulator length to saturation is in the range from about nine (hard x-ray SASE
FELs) to eleven (visible range SASE FELs) field gain lengths
\cite{coherence-oc}. Situation with the transverse coherence is favorable when
the relative separation of the field gain between fundamental and higher modes
exceeds 25-30 \%. In this case the maximum degree of transverse coherence can
exceed the value of 90 \% \cite{coherence-oc,trcoh-oc}. Further development of
the amplification process in the nonlinear stage leads to visible degradation
of the coherence properties.

Separation of the gain of the FEL radiation modes mainly depends on the value
of the diffraction parameter. Increase of the value of the diffraction parameter results in
less relative separation of the gain of the modes. In this case we deal with
the mode degeneration \cite{fs2r-1993,book}. Since the number of gain
lengths to saturation is limited, the contribution of the higher spatial
modes to the total power grows with the value of the diffraction parameter,
and the transverse coherence degrade. Parameter range of large diffraction
parameter values is typical for SASE FELs operating in the hard x-ray
wavelength range \cite{euro-xfel-tdr,lcls,sacla,swiss-fel,pal-xfel}. It is also worth noticing that
a spread of longitudinal velocities (due to energy spread and emittance) helps to suppress high order
modes thus improving transverse coherence properties. This consideration suggests that a tight focusing
of the electron beam in the undulator can be important for reaching a good coherence due to a reduction of the
diffraction parameter and an increase of the velocity spread.

In this paper we perform thorough analysis of the coherence properties of the
radiation from FLASH free electron laser. We found that the degree of
transverse coherence of the radiation from FLASH is visible less than unity
in the post-saturation regime. Moreover, we find that the pointing stability of the FEL beam
suffers from not sufficient mode selection of higher spatial radiation modes which
happens due to large values of the diffraction parameter. Our analysis shows
that operation with a stronger focusing of the electron beam and a lower peak
current would allow to improve degree of transverse coherence and the pointing stability.

\section{Analysis of the radiation modes}

We consider axisymmetric model of the electron beam. It is assumed that
transverse distribution function of the electron beam is Gaussian, so rms
transverse size of matched beam is $\sigma = \sqrt{\epsilon \beta }$, where
$\epsilon $ is rms beam emittance and $\beta $ is the beta-function.
In the framework of the three-dimensional theory the
operation of a short-wavelength FEL amplifier is described by the following parameters: the
diffraction parameter $B$, the energy spread parameter $\hat{\Lambda
}^{2}_{\mathrm{T}}$, the betatron motion parameter $\hat{k}_{\beta }$ and
detuning parameter $\hat{C}$ \cite{book,ssy-eigen-bet}:

\begin{eqnarray}
B & =& 2 \Gamma \sigma^2 \omega/c \ , \qquad \hat{C} = C/\Gamma \ ,
\nonumber \\
\hat{k}_{\beta} & = & 1/(\beta \Gamma ) \ ,
\qquad \hat{\Lambda }^{2}_{\mathrm{T}}  =
(\sigma _{\mathrm{E}}/E)^2/\rho^2  \ ,
\label{eq:reduced-parameters}
\end{eqnarray}

\begin{figure}[b]

\includegraphics[width=0.5\textwidth]{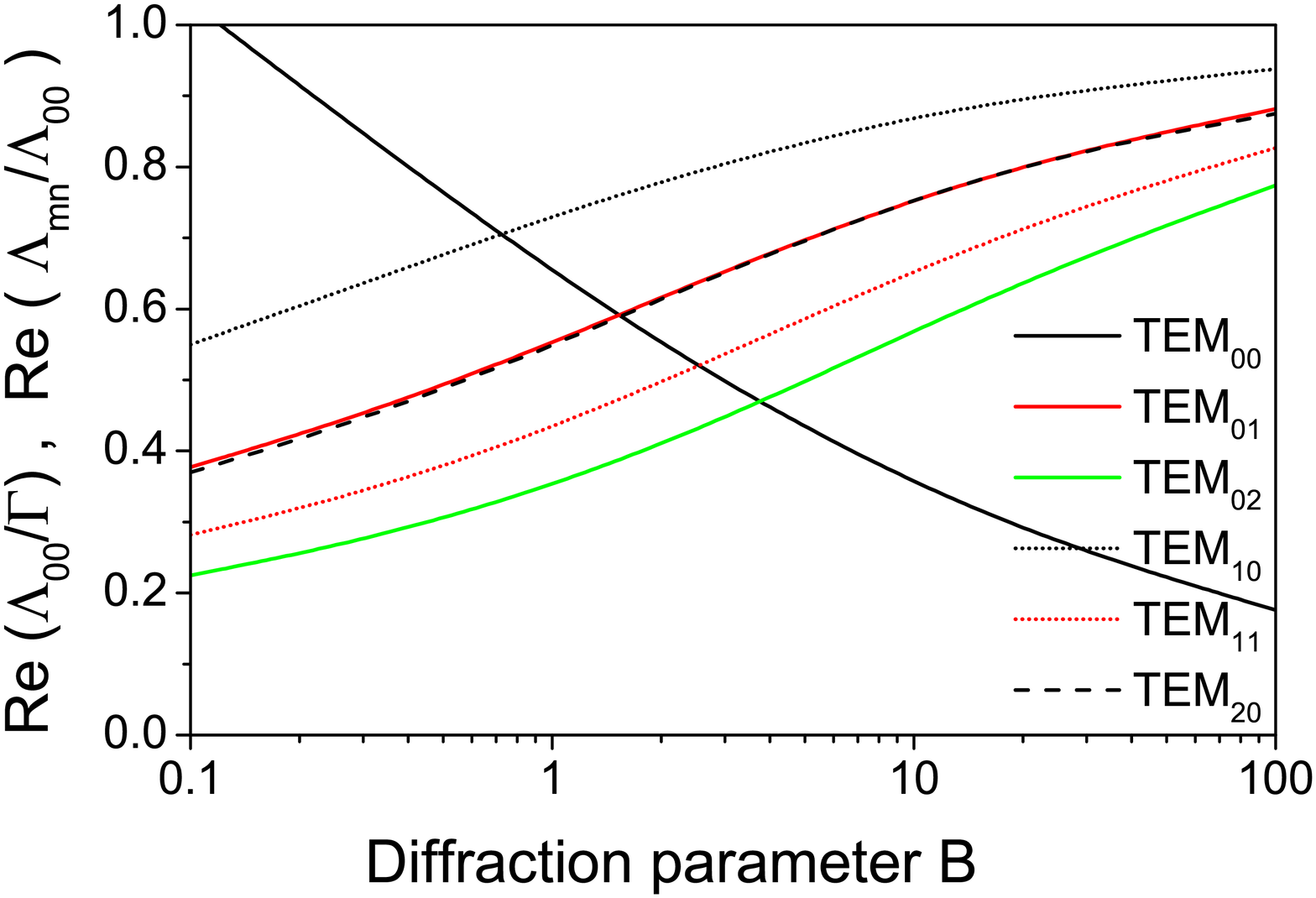}

\caption{
Ratio of the maximum gain of the higher modes to the maximum gain of the
fundamental mode $\re (\Lambda _{mn})/\re (\Lambda _{00})$ versus diffraction
parameter $B$. The energy spread parameter is $\hat{\Lambda }^{2}_{\mathrm{T}}
\to 0$, and the betatron motion parameter is $\hat{k}_{\beta } \to 0$. Color
codes refer to the radial index of the mode: 0 - black, 1 - red, 2 - green.
Line type codes refer to the azimuthal index of the mode: 0 - solid line, 1 -
dotted line, 2 - dashed line. Black solid line shows the gain of the fundamental
mode $\re (\Lambda _{00})/ \Gamma $.
}

\label{fig:relb10}
\end{figure}

\begin{figure}[tb]

\includegraphics[width=0.5\textwidth]{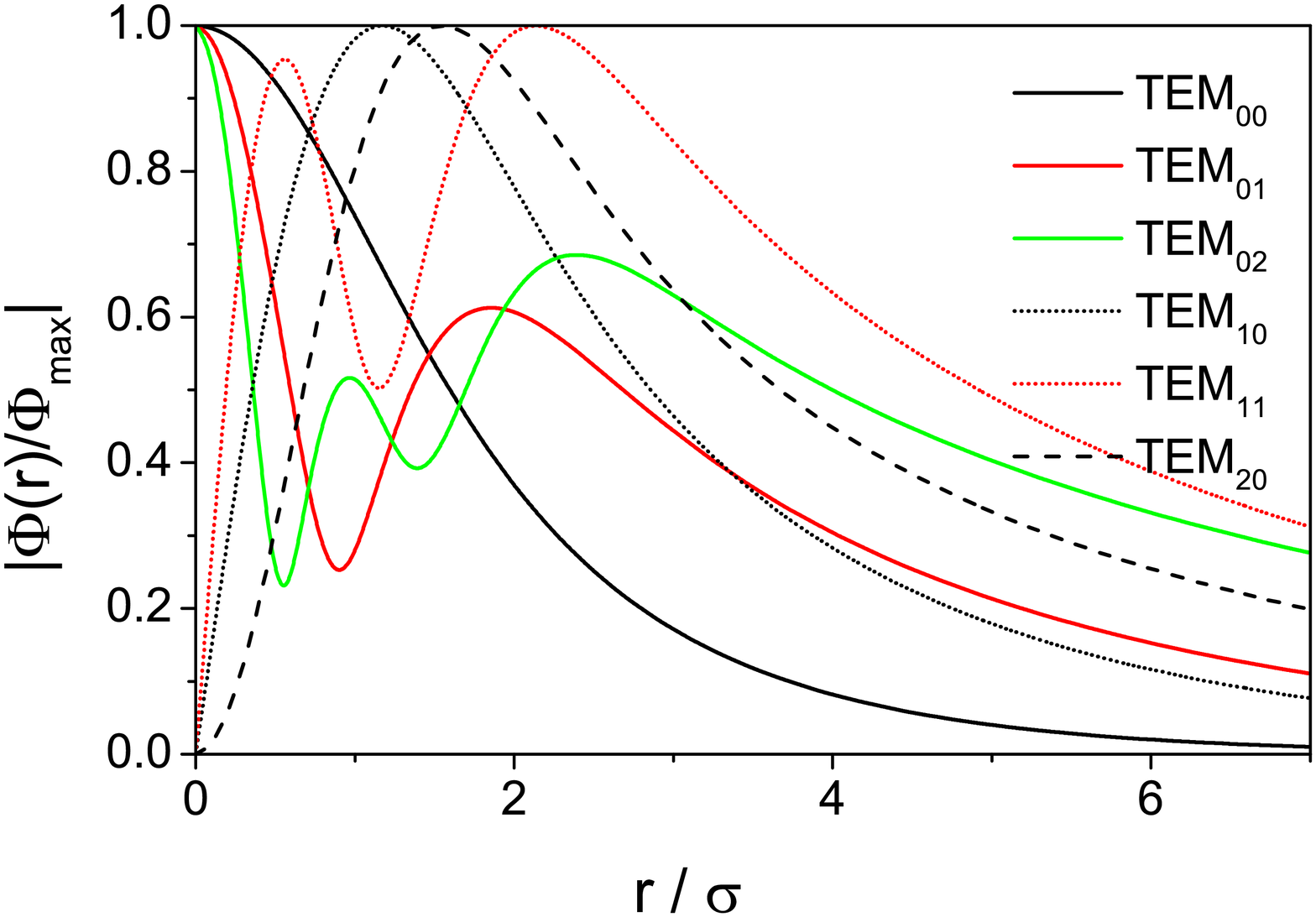}
\includegraphics[width=0.5\textwidth]{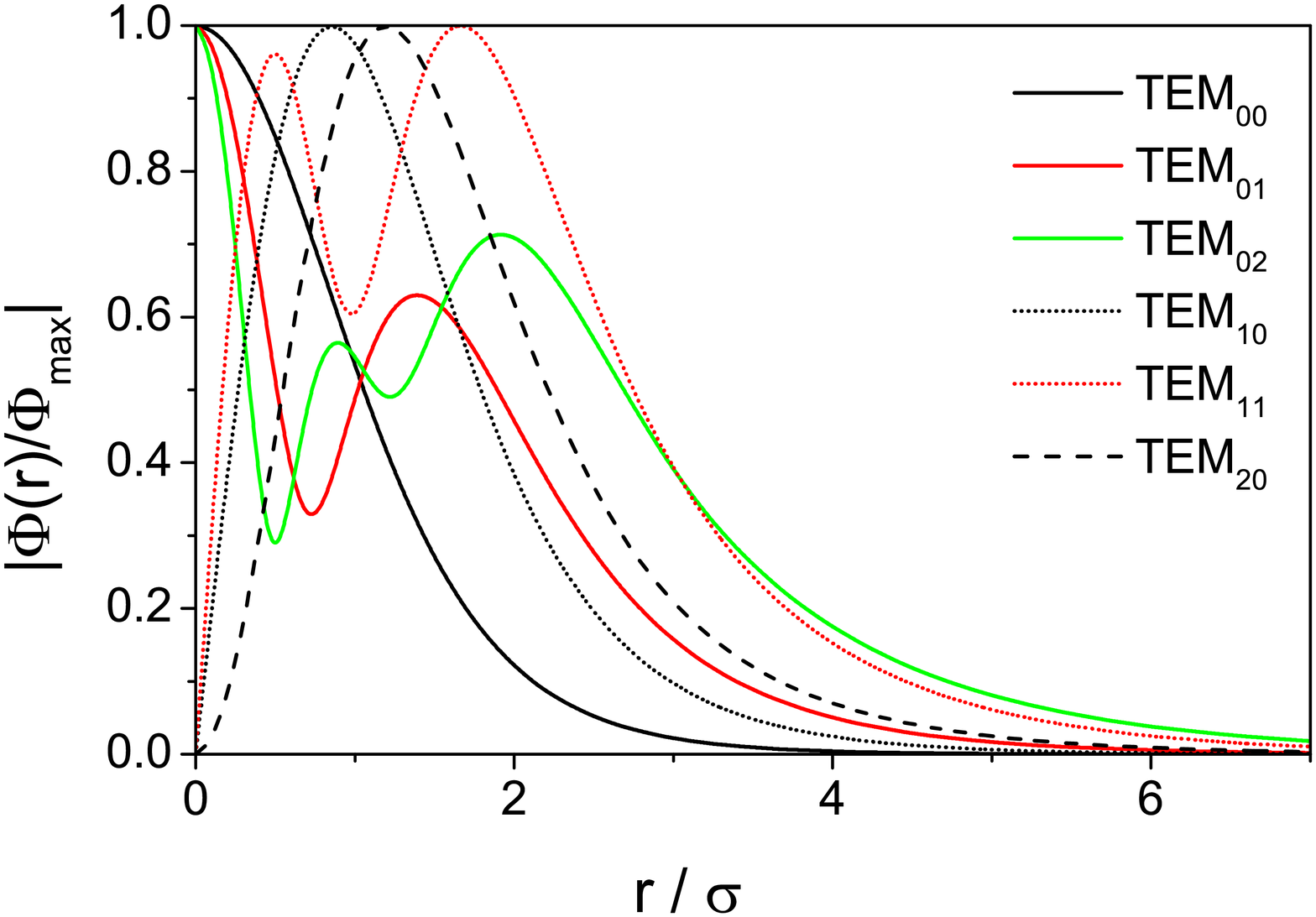}

\caption{
Amplitude of the eigenfunctions of the FEL radiation modes, $|\Phi
_{mn}(r)|/|\Phi _{\max}|$.
Left and right plot correspond to the diffraction parameter $B =1$ and $B = 10$,
respectively.
The detuning corresponds to the maximum of the gain.
The energy spread parameter is $\hat{\Lambda }^{2}_{\mathrm{T}}
\to 0$, and the betatron motion parameter is $\hat{k}_{\beta } \to 0$. Color
codes refer to the radial index of the mode: 0 - black, 1 - red, 2 - green.
Line type codes refer to the azimuthal index of the mode: 0 - solid line, 1 -
dotted line, 2 - dashed line.
}

\label{fig:ntemn}
\end{figure}

\begin{figure}[tb]

\includegraphics[width=0.5\textwidth]{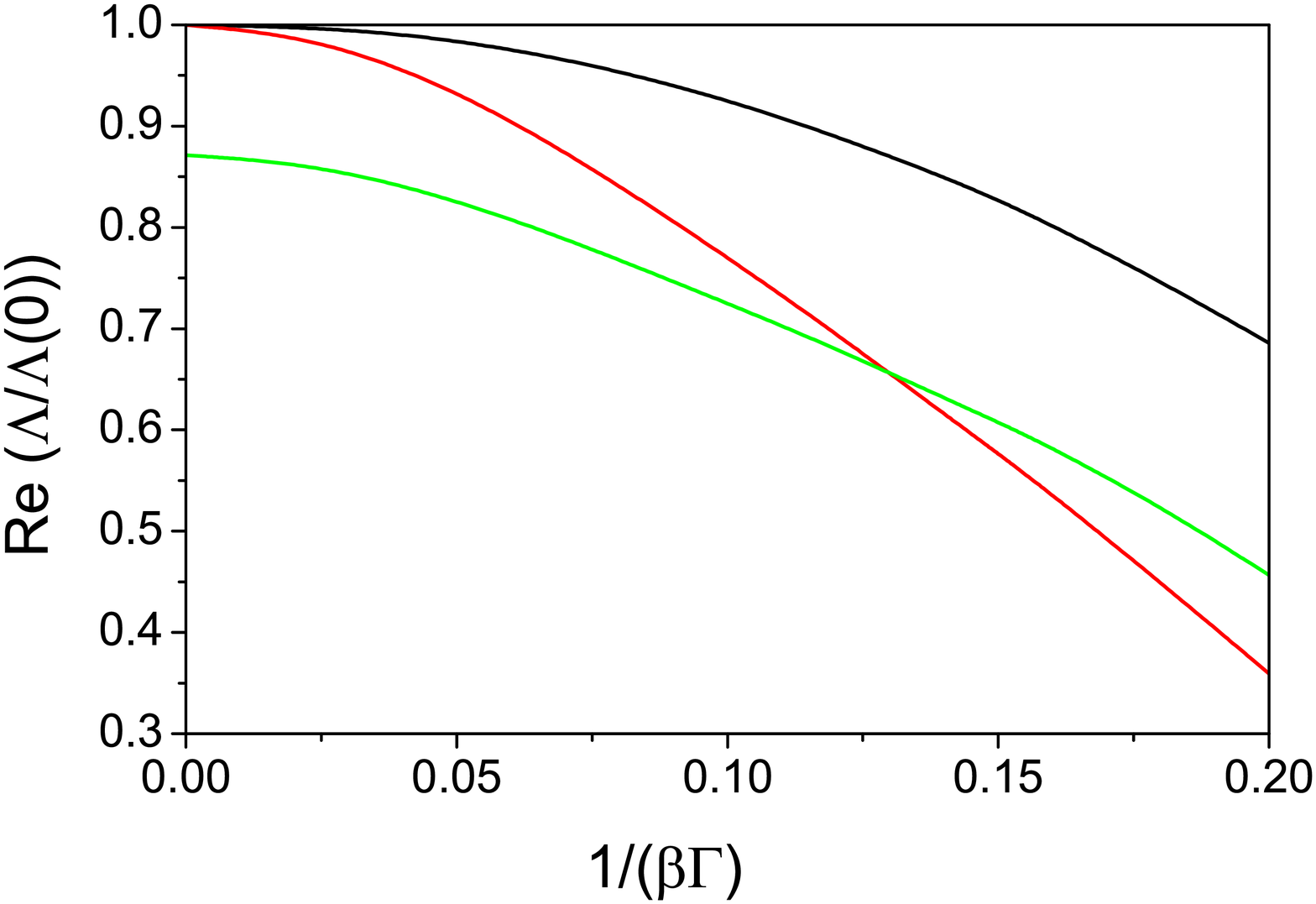}

\caption{
Dependence of the gain of TEM$_{00}$ mode (black curve) and TEM$_{10}$
mode (red curve) on the betatron motion parameter $\hat{k}_{\beta
} = 1/(\beta \Gamma )$.
The values are normalized to those at $\hat{k}_{\beta
} \to 0$. Green curve shows the ratio of the gain of TEM$_{10}$ mode to the
gain of TEM$_{00}$ mode.
The diffraction parameter is $B = 10$.
The energy spread parameter is $\hat{\Lambda }^{2}_{\mathrm{T}}
\to 0$.
}

\label{fig:betgamb10}
\end{figure}

\begin{figure}[tb]

\includegraphics[width=0.5\textwidth]{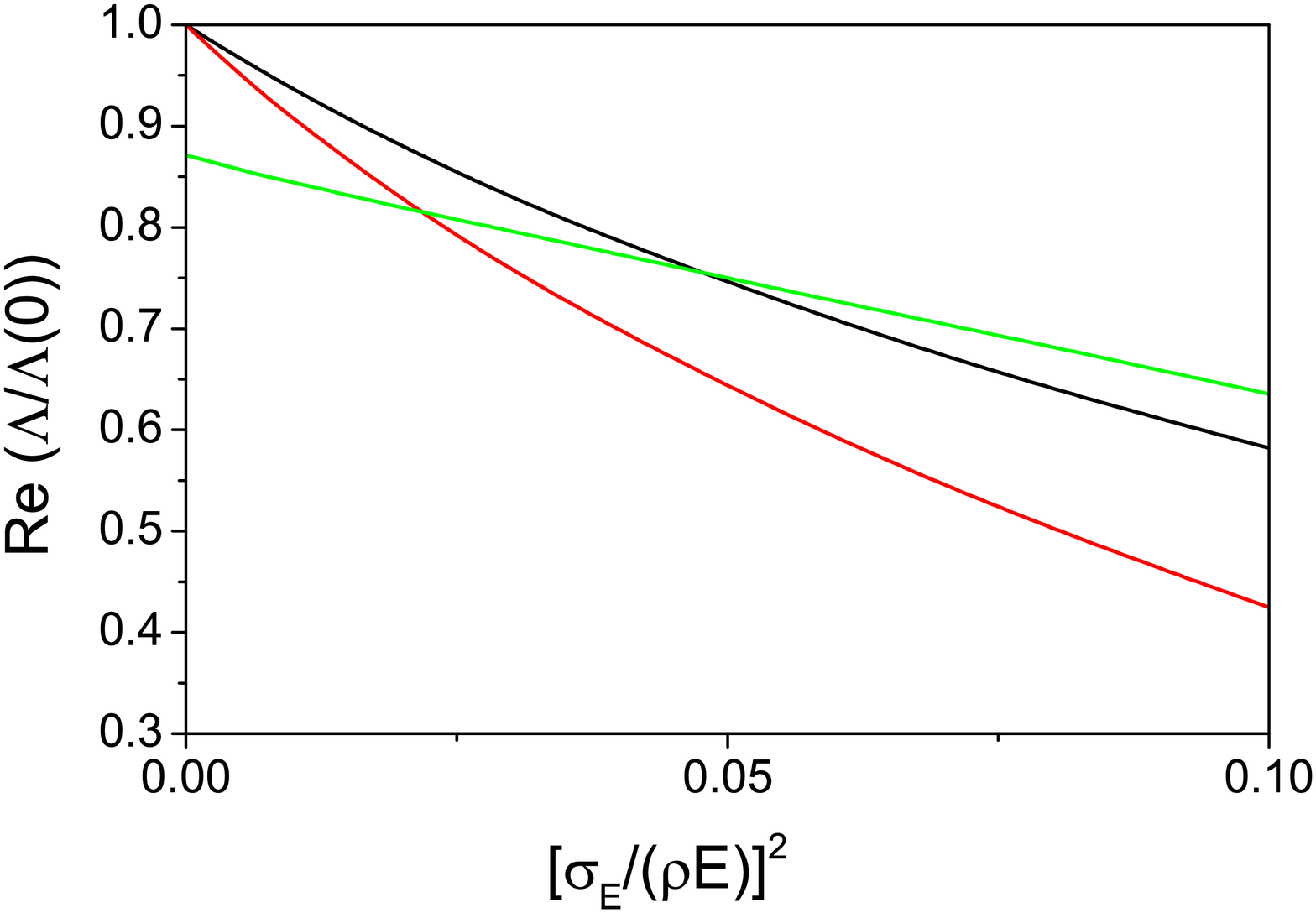}

\caption{
Dependence of the gain of TEM$_{00}$ mode (black curve) and TEM$_{10}$
mode (red curve) on the energy spread parameter $\hat{\Lambda }^{2}_{\mathrm{T}}$.
The values are normalized to those at $\hat{\Lambda }^{2}_{\mathrm{T}} \to 0$.
Green curve shows the ratio of the gain of TEM$_{10}$ mode to the
gain of TEM$_{00}$ mode.
The diffraction parameter is $B = 10$.
The betatron oscillation parameter is $\hat{k}_{\beta } \to 0$.
}

\label{fig:lt2b10}
\end{figure}

\noindent where $E = \gamma mc^2$ is the energy of electron, $\gamma $ is
relativistic factor, $\Gamma = \left[ I \omega^2 \theta_{\mathrm{s}}^2
A_{\mathrm{JJ}}^2/ (I_{\mathrm{A}} c^2 \gamma_{\mathrm{z}}^2 \gamma )
\right]^{1/2}$ is the gain parameter, $\rho = c\gamma ^{2}_{\mathrm{z}}\Gamma
/\omega $ is the efficiency parameter, and $C = 2\pi/\lambda _{\mathrm{w}} -
\omega /(2c\gamma ^{2}_{z})$ is the detuning of the electron with the nominal
energy ${\cal E}_{0}$. Note that the efficiency parameter $\rho $ entering
equations of three dimensional theory relates to the one-dimensional parameter
$\rho _{\mathrm{1D}}$ as $\rho _{\mathrm{1D}} = \rho / B^{1/3}$
\cite{book,boni-rho}. The following notations are used here: $I$ is the beam
current, $\omega = 2\pi c/\lambda$ is the frequency of the electromagnetic
wave, $\lambda _{\mathrm{w}}$ is undulator period,
$\theta_{\mathrm{s}}=K/\gamma$, $K$ is the rms undulator parameter, $\gamma
^{-2}_{z} = \gamma ^{-2}+ \theta ^{2}_{\mathrm{s}}$, $I_{\mathrm{A}} = mc^3/e = $
17 kA is the Alfven current, $A_{\mathrm{JJ}} = 1$ for helical undulator and
$A_{\mathrm{JJ}} = J_0(K^2/2(1+K^2)) - J_1(K^2/2(1+K^2))$ for planar undulator.
$J_0$ and $J_1$ are the Bessel functions of the first kind. The energy spread
is assumed to be Gaussian with rms deviation $\sigma _{\mathrm{E}}$.

Amplification process in SASE FEL starts from the shot noise in the electron
beam. At the initial stage of amplification coherence properties are poor, and
the radiation consists of a large number of transverse and longitudinal modes
\cite{trcoh-oc,moore-modes,fs2r-1993,xie-modes,ssy-eigen-bet,book,kim-inval-1,wang-yu,yu-krinsky}:

\begin{equation}
\tilde{E} =
\sum \limits _{m,n}\int \mathrm{d}\omega
A_{mn}(\omega , z) \Phi_{mn}(r,\omega )
\exp [ \Lambda _{mn}(\omega )z + im\phi + i\omega (z/c-t)] \ .
\label{eq:modes}
\end{equation}

\noindent Each mode is characterized by the eigenvalue $\Lambda _{mn}(\omega )$
and the field distribution eigenfunction $\Phi _{mn}(r,\omega )$. Real part of
the eigenvalue $\re (\Lambda _{mn}(\omega ))$ is referred as the field gain.
The field gain length is $L_g = 1/\re (\Lambda _{mn}(\omega ))$. Eigenvalues
and eigenfunctions are the solutions of the eigenvalue equation
\cite{xie-modes,ssy-eigen-bet}. Each eigenvalue has a maximum at a certain frequency (or, at
a certain detuning), so that the detunig for each mode is chosen automatically in the case of a SASE FEL (in
contrast with seeded FELs where the detuning can be set to any value). Thus, we will in fact deal
with the three dimensionless parameters:
$B$, $\hat{k}_{\beta}$, and $\hat{\Lambda }^{2}_{\mathrm{T}}$.

Let us look closer at the properties of the
radiation modes. The gains for several modes is depicted in
Fig.~\ref{fig:n1n0b} as functions of the diffraction parameter. The values for
the gain correspond to the maximum of the scan over the detuning parameter
$\hat{C}$. The curve for TEM$_{00}$ mode shows the values of normalized gain
$\re (\Lambda _{00}/\Gamma )$. Curves for the higher spatial modes present the
ratio of the gain of the mode to the gain of the fundamental mode, $\re
(\Lambda _{mn}/\Lambda _{00})$. Sorting of the modes by the gain results in the
following ranking: TEM$_{00}$, TEM$_{10}$, TEM$_{01}$, TEM$_{20}$, TEM$_{11}$,
TEM$_{02}$. The gain of the fundamental TEM$_{00}$ mode is always above the
gain of higher order spatial modes. The difference in the gain between the
fundamental TEM$_{00}$ mode and higher spatial modes is pronouncing for small
values of the diffraction parameter $B \lesssim 1$. The gain of higher spatial
modes approaches asymptotically the gain of the fundamental mode for large
values of the diffraction parameter. In other words, the effect of the mode
degeneration takes place. Its origin can be understood with the qualitative
analysis of the eigenfunctions (distribution of the radiation field in the near
zone). Figure~\ref{fig:ntemn} shows eigenfunctions of the FEL radiation modes
for two values of the diffraction parameter, $B = 1$ and $B = 10$. We
observe that for small values of the diffraction parameter the field of the
higher spatial modes spans far away from the core of the electron beam while
the fundamental TEM$_{00}$ mode is more confined. This feature provides higher
coupling factor of the radiation with the electron beam and higher gain. For
large values of the diffraction parameter all radiation modes shrink to the
beam axis which results in equalizing of coupling factors and of the gain.
Asymptotically, the eigenvalues of all modes tends to the one dimensional
asymptote as \cite{coherence-anal-oc}:

\begin{equation}
\Lambda_{mn}/\Gamma \simeq \frac{\sqrt{3}+\I}{2B^{1/3}} -
\frac{(1+\I\sqrt{3})(1+n+2m)}{3\sqrt{2}B^{2/3}}
\label{growth-cold}
\end{equation}

For a SASE FEL, the undulator length to saturation is in the range from about nine
(hard x-ray range) to eleven (visible range) field gain lengths
\cite{coherence-oc,coherence-njp,coh-fel2012}. The mode selection process stops
at the onset of nonlinear regime, about two field gain length before
the saturation. Let us make simple estimation for the value of the diffraction
parameter $B = 10$ and cold electron beam, $\hat{\Lambda }^{2}_{\mathrm{T}} \to
0$, and $\hat{k}_{\beta } \to 0$. We get from Fig.~\ref{fig:n1n0b} that the
ratio of the gain $\re (\Lambda _{10} / \Lambda _{00})$ is equal to 0.87. With
an assumption of similar values of coupling factors, we find that the ratio of
the field amplitudes at the onset of the nonlinear regime is about of
factor of 3 only. An estimate for the contribution of the higher spatial modes
to the total power is about 10 \%. Another numerical example for $B = 1$ gives
the ratio $\re (\Lambda _{10} / \Lambda _{00}) = 0.73$, and the ratio of field
amplitudes exceeds a factor of 10. Thus, an excellent transverse coherence of the
radiation is not expected for SASE FEL with diffraction parameter $B \gtrsim 10$
and a small velocity spread in the electron beam.

Longitudinal velocity spread due to the
energy spread and emittance serves as a tool for selective suppression of the
gain of the higher spatial modes \cite{fs2r-1993,book}.
Figures~\ref{fig:betgamb10} and  \ref{fig:lt2b10} show the dependence of the
gain of TEM$_{00}$ and TEM$_{10}$ modes on the betatron motion parameter and
the energy spread parameter. We see that with the fixed value of the diffraction
parameter, the mode degeneration effect can be relaxed at the price of the gain
reduction.

The betatron motion can influence the gain of different modes (and, therefore, transverse coherence properties)
via two different mechanisms. First, the particles move across the beam thus transferring the information between
different points in the beam cross-section. Second, as it was already mentioned, there is a spread of longitudinal
velocities that has a similar effect as the energy spread (and it is usually more important than the first one).
One can introduce a combination of parameters $B$ and
$\hat{k}_{\beta }$ that can to some extent be similar with the energy spread parameter:

\begin{equation}
(\hat{\Lambda }^{2}_{\mathrm{T}})_{eff} =
B^2 \hat{k}_{\beta }^4
\label{es-eff}
\end{equation}

Finally, let us note that the situation with transverse coherence is favorable when relative separation of
the gain between the fundamental and higher spatial modes is more than 25-30 \%.
In this case the degree of transverse coherence
can reach values above 90\% in the
end of the high gain linear regime \cite{coherence-anal-oc,trcoh-oc}. Further
development of the amplification process in the nonlinear stage leads to a
significant degradation of the spatial and of the temporal coherence
\cite{coherence-oc,coherence-njp,coh-fel2012}.

\section{Parameter space of FLASH}

In the present experimental situation, many parameters of the electron beam at
FLASH depend on practical tuning of the machine. Analysis of measurements and
numerical simulations shows that depending on tuning of the machine emittance
may change from about 1 to about 1.5 mm-mrad. Tuning at small charges may allow
to reach smaller values of the emittance down to 0.5 mm-mrad. Peak current may
change in the range from 1 kA to 2 kA depending on the tuning of the beam
formation system. An estimate for the local energy spread is $\sigma
_{\mathrm{E}}$ [MeV]  $\simeq 0.1 \times I $ [kA]. The average beta
function in the undulator is about 10 meters.

\begin{figure}[tb]

\includegraphics[width=0.5\textwidth]{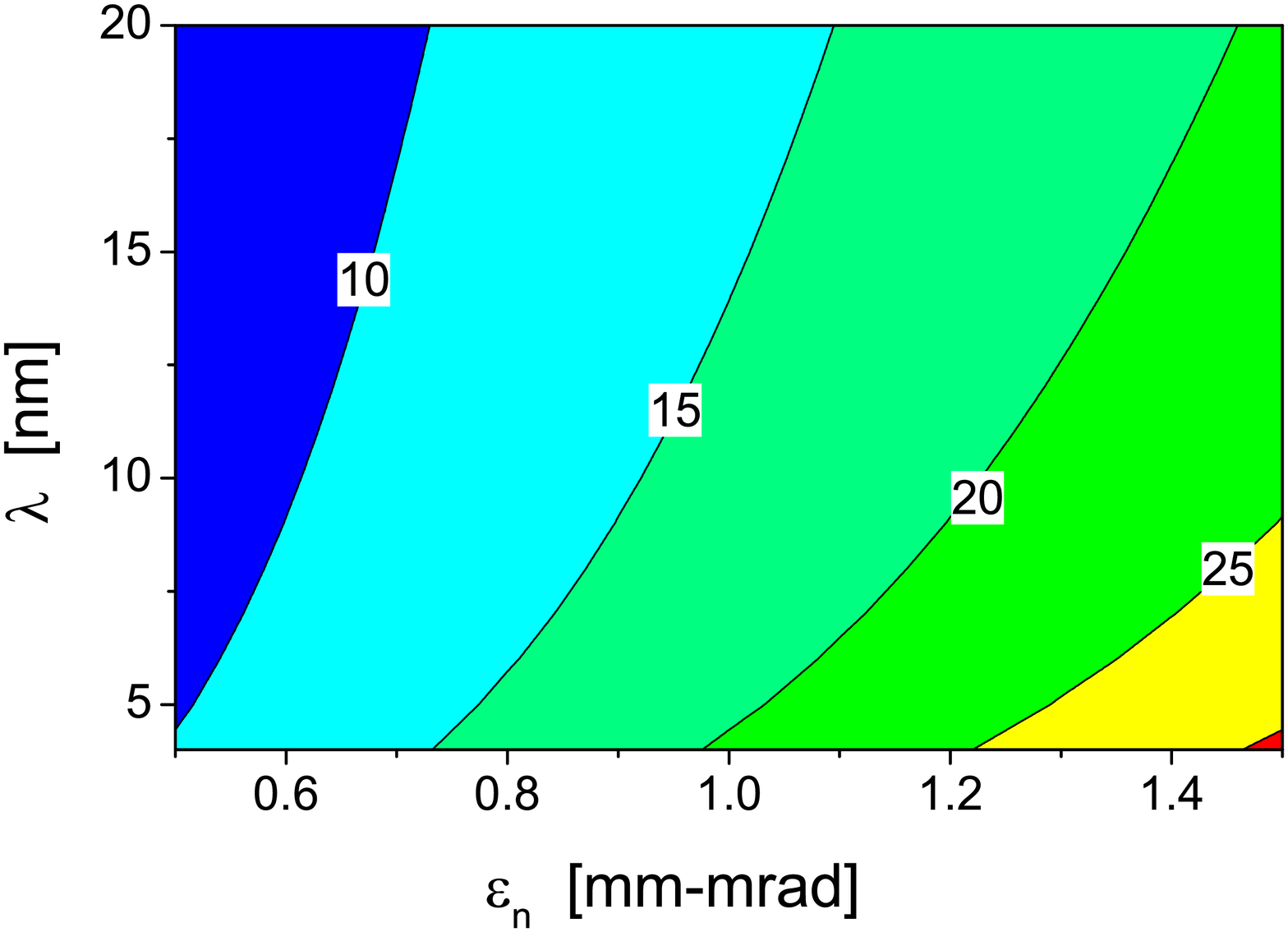}
\includegraphics[width=0.5\textwidth]{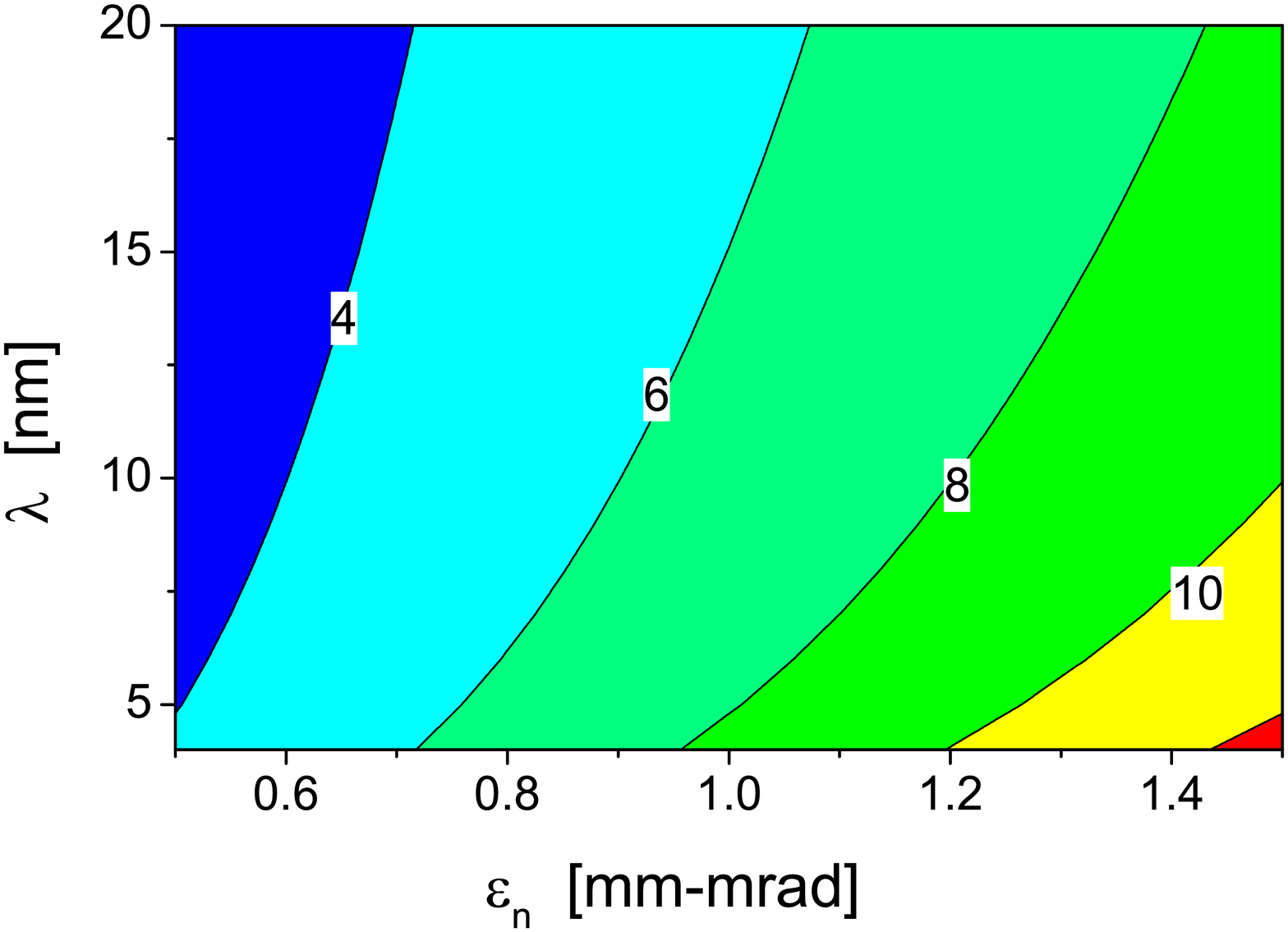}

\caption{
Contour plot for the value of the diffraction parameter $B$ versus normalized
emittance and radiation wavelength.
Left plot: beam current is 1.5 kA, beta function is 10 m.
Right plot: beam current is 1 kA, beta function is 5 m.
}

\label{fig:bemlamb10i15}

\end{figure}

\begin{figure}[tb]

\includegraphics[width=0.5\textwidth]{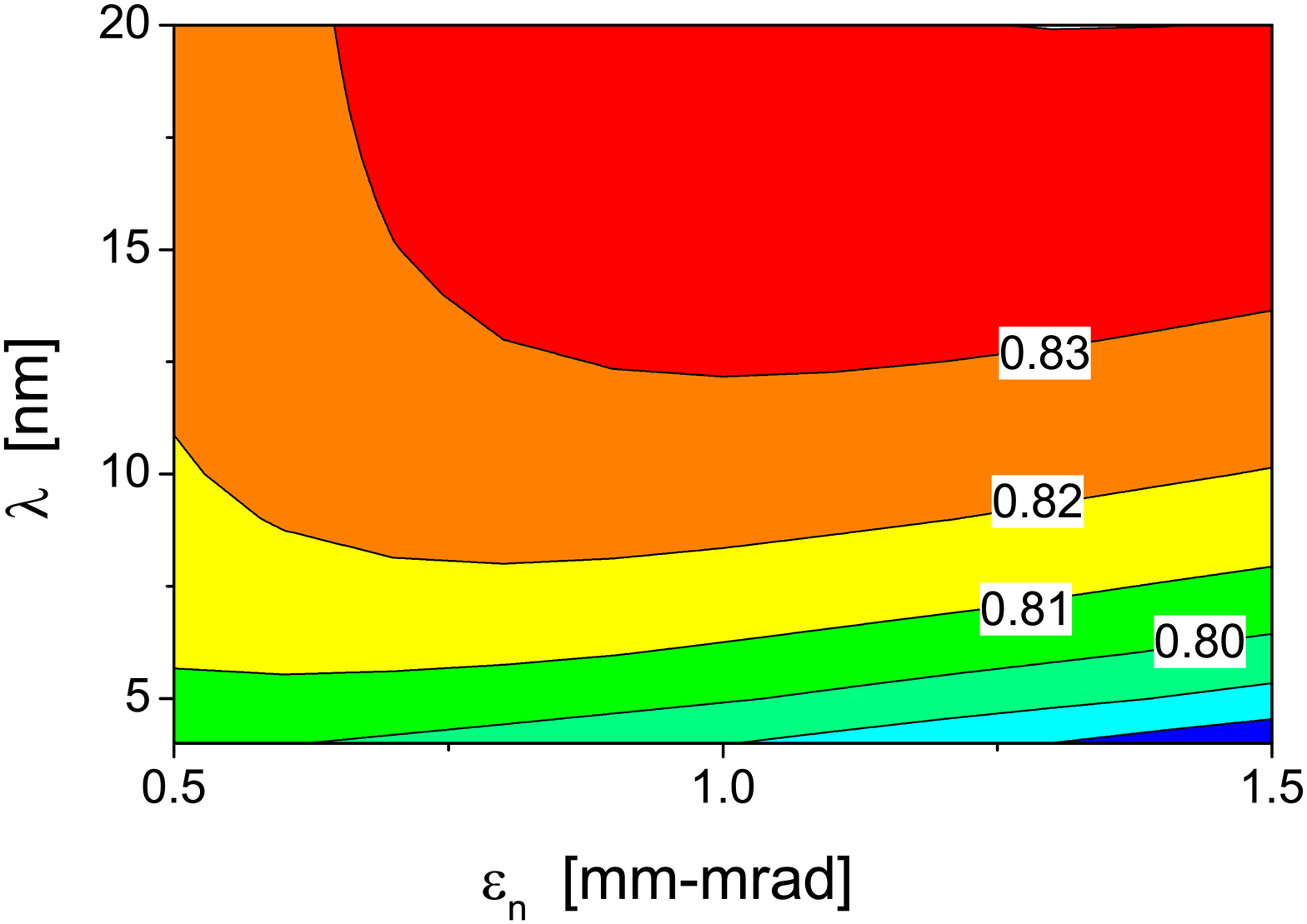}
\includegraphics[width=0.5\textwidth]{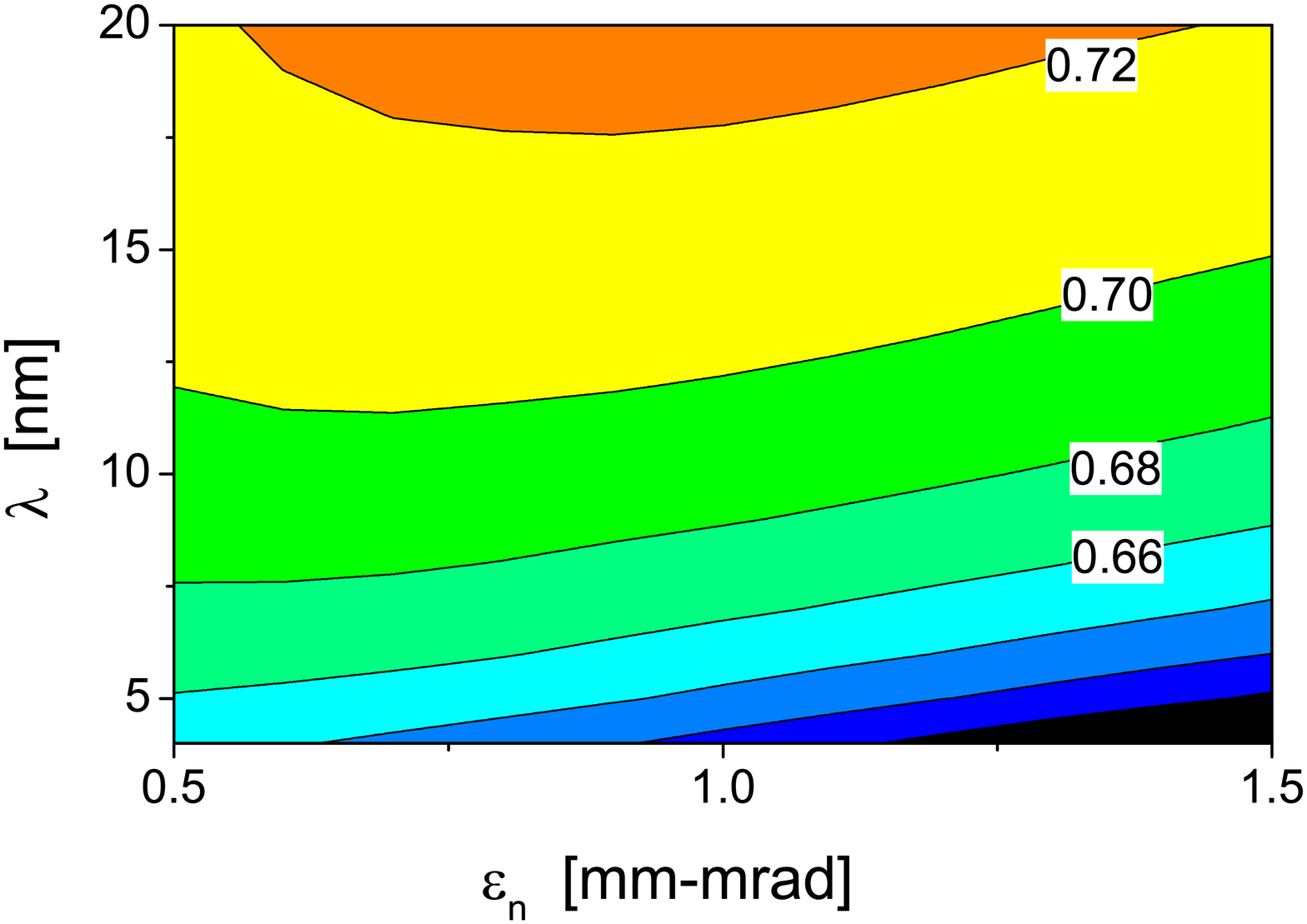}

\caption{
Contour plot of the ratio of the maximum field gain of TEM$_{10}$ to the field
gain of the ground TEM$_{00}$ mode versus radiation wavelength and emittance.
Left plot: beam current is 1.5 kA, beta function is 10 m.
Right plot: beam current is 1 kA, beta function is 5 m.
}

\label{fig:n1n0b}
\end{figure}

Let us choose the reference working point with the radiation wavelength 8 nm,
rms normalized emittance 1 mm-mrad and beam current 1.5 kA. Parameters of the
problem for this reference point are: the diffraction parameter is $B = 17.2$,
the energy spread parameter $\hat{\Lambda }^{2}_{\mathrm{T}}  = 1.7 \times
10^{-3}$, betatron motion parameter $\hat{k}_{\beta} = 5.3 \times
10^{-2}$.

\noindent Then the reduced parameters at other working points can be easily
recalculated using the scaling:

$$
\hat{k}_{\beta} \propto \frac{1}{\beta  I^{1/2} \lambda ^{1/4}} \qquad
\hat{\Lambda }^{2}_{\mathrm{T}}  \propto I \lambda ^{1/2} \qquad
B \propto \frac{\epsilon _n \beta I^{1/2}}{\lambda ^{1/4}} \ .
$$

\noindent The effective contribution of the emittance to the longitudinal velocity spread (\ref{es-eff})
scales as follows

$$
(\hat{\Lambda }^{2}_{\mathrm{T}})_{eff}  \propto
\frac{\epsilon _n^2}{ \beta^2 I \lambda ^{3/2}}
$$

\noindent and equals $2.3 \times 10^{-3}$ at the considered working point.

Analyzing these simple dependencies in terms of their effect on mode separation, we can state that

\begin{itemize}

\item Dependencies on the wavelength are relatively weak (except for $(\hat{\Lambda }^{2}_{\mathrm{T}})_{eff}$), i.e.
one should not expect a significantly better transverse coherence at longer wavelengths. Moreover, mode separation can even be
somewhat improved at shorter wavelengths due to a significant increase in $(\hat{\Lambda }^{2}_{\mathrm{T}})_{eff}$.

\item Reduction of the peak current (by going to a weaker bunch compression) would lead to an improvement of
mode separation (even though the energy spread parameter would smaller). Obviously, the peak power at the saturation
would be reduced.

\item Dependence on the normalized emittance is expected to be weak because of the two competing effects. Mode
separation due to a change of the diffraction parameter can be to a large extent compensated by a change of the longitudinal
velocity spread. As we will see below, this happens indeed in the considered parameter range.

\item Reduction of the beta-function would be the most favorable change because it would reduce the diffraction parameter, and
increase the velocity spread at the same time.
Unfortunately, there are technical arguments not supporting such a change
in the FLASH undulator.
Nevertheless, for illustration we will also present some results for the beta-function equal to 5 m.

\end{itemize}

\noindent Contour plot for the value of the diffraction parameter $B$ for the
value of beta function of 10 m and the value of beam current 1.5 kA is
presented in Fig.~\ref{fig:bemlamb10i15}. We see that the value of the
diffraction parameter is $B \gtrsim 10$ in the whole parameter space.
Figure~\ref{fig:n1n0b} shows the ratio of the field gain
$\re (\Lambda _{10}(\omega ))$ to the value of the field gain $\re
(\Lambda _{00}(\omega ))$ of the fundamental mode. We see that this ratio is
above 0.8 in the whole range of parameters, and we can expect significant
contribution of the first azimuthal mode to the total radiation power. We can also notice relatively weak dependencies
on the emittance and on the wavelength. However, reduction of the current and of the
beta-function help to increase the mode separation up to a desirable level as one can see from the right plot
in Fig.~\ref{fig:n1n0b}

\section{Simulation procedure}

Simulations have been performed with three-dimensional, time-dependent FEL
simulation code \cite{fast}. The result of each simulation run contains an
array of complex amplitudes $\tilde{E}$ for electromagnetic fields on a
three-dimensional mesh. At the next stage of the numerical experiment the data
arrays are handled with postprocessor codes to calculate different
characteristics of the radiation. Simulations of the statistical properties
have been performed for the case of a long bunch with uniform axial profile of
the beam current.

\begin{figure}[tb]

\includegraphics[width=0.5\textwidth]{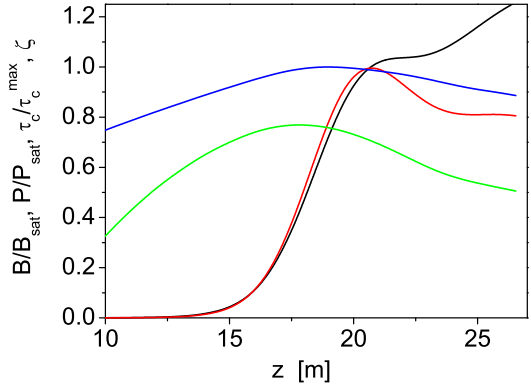}

\caption{
Evolution of the radiation power (black curve), coherence time (blue curve),
degree of transverse coherence (green curve), and brilliance (red curve) along
the undulator. Brilliance and radiation power are normalized to saturation
values. Coherence time is normalized to maximum value of 5.5 fs.
Radiation wavelength is 8 nm.
Beta function is 10 m.
Beam current is 1.5 kA.
rms normalized emittance is 1 mm-mrad.
}
\label{fig:dctpz1510}
\end{figure}

The first-order time correlation function $g_1(t,t')$ and transverse
correlation function $\gamma_1 (\vec{r}_{\perp}, \vec{r}\prime _{\perp}, z, t)$
are defined as

\begin{displaymath}
g_1(\vec{r},t-t')  =
\frac{\langle \tilde{E}(\vec{r},t)\tilde{E}^*(\vec{r},t')\rangle }
{\left[\langle \mid\tilde{E}(\vec{r},t)\mid^2\rangle
\langle \mid\tilde{E}(\vec{r},t')\mid^2\rangle \right]^{1/2}} \ ,
\end{displaymath}

\begin{displaymath}
\gamma_1 (\vec{r}_{\perp}, \vec{r}\prime _{\perp}, z, t) = \frac{
\langle \tilde{E} (\vec{r}_{\perp}, z, t)
\tilde{E}^{*} (\vec{r}\prime _{\perp}, z, t) \rangle }
{ \left[ \langle |\tilde{E} (\vec{r}_{\perp}, z, t) |^2 \rangle
\langle |\tilde{E} (\vec{r}\prime _{\perp}, z, t) |^2 \rangle \right]^{1/2}}
\ ,
\end{displaymath}

\noindent where $\tilde{E}$ is the slowly varying amplitude of the amplified
wave. For a stationary random process the coherence time and the degree of
transverse are defined as \cite{coherence-oc}:

\begin{eqnarray}
&\mbox{}&
\zeta =
\frac
{
\int |\gamma _1 (\vec{r}_{\perp},\vec{r}\prime _{\perp})|^{2}
I(\vec{r}_{\perp})
I(\vec{r}\prime _{\perp})
\D\vec{r}_{\perp}
\D\vec{r}\prime _{\perp}
}
{
[\int I(\vec{r}_{\perp}) \D\vec{r}_{\perp}]^{2}
} \ ,
\nonumber \\
&\mbox{}&
\tau_{\mathrm{c}} = \int \limits^{\infty}_{-\infty} | g_1(\tau) |^2 \D\tau \ ,
\label{eq:def-degcoh}
\end{eqnarray}

\noindent where $I(\vec{r}_{\perp}) =
\langle |\tilde{E} (\vec{r}_{\perp}) |^2 \rangle $.
The first order time correlation function, $g_1(t,t')$, is
calculated in accordance with the definition:

\noindent Brilliance of the radiation is proportional to the product of the
radiation power, coherence time, and of the degree of transverse coherence.
Evolution of general characteristics of SASE FEL along the undulator is
illustrated in
Figure~\ref{fig:dctpz1510}. If one traces evolution of the brilliance of the
radiation along the undulator length, there is always the point, which we
define as the saturation point, where the brilliance reaches maximum value
\cite{coherence-oc}.

\section{Results of numerical simulations}

We illustrate general characteristics of FLASH with specific numerical example
for FLASH operating at the wavelength of 8 nm, peak current 1.5 kA, and rms
normalized emittance 1 mm-mrad.

\subsection{Radiation power}

\begin{figure}[tb]

\includegraphics[width=0.5\textwidth]{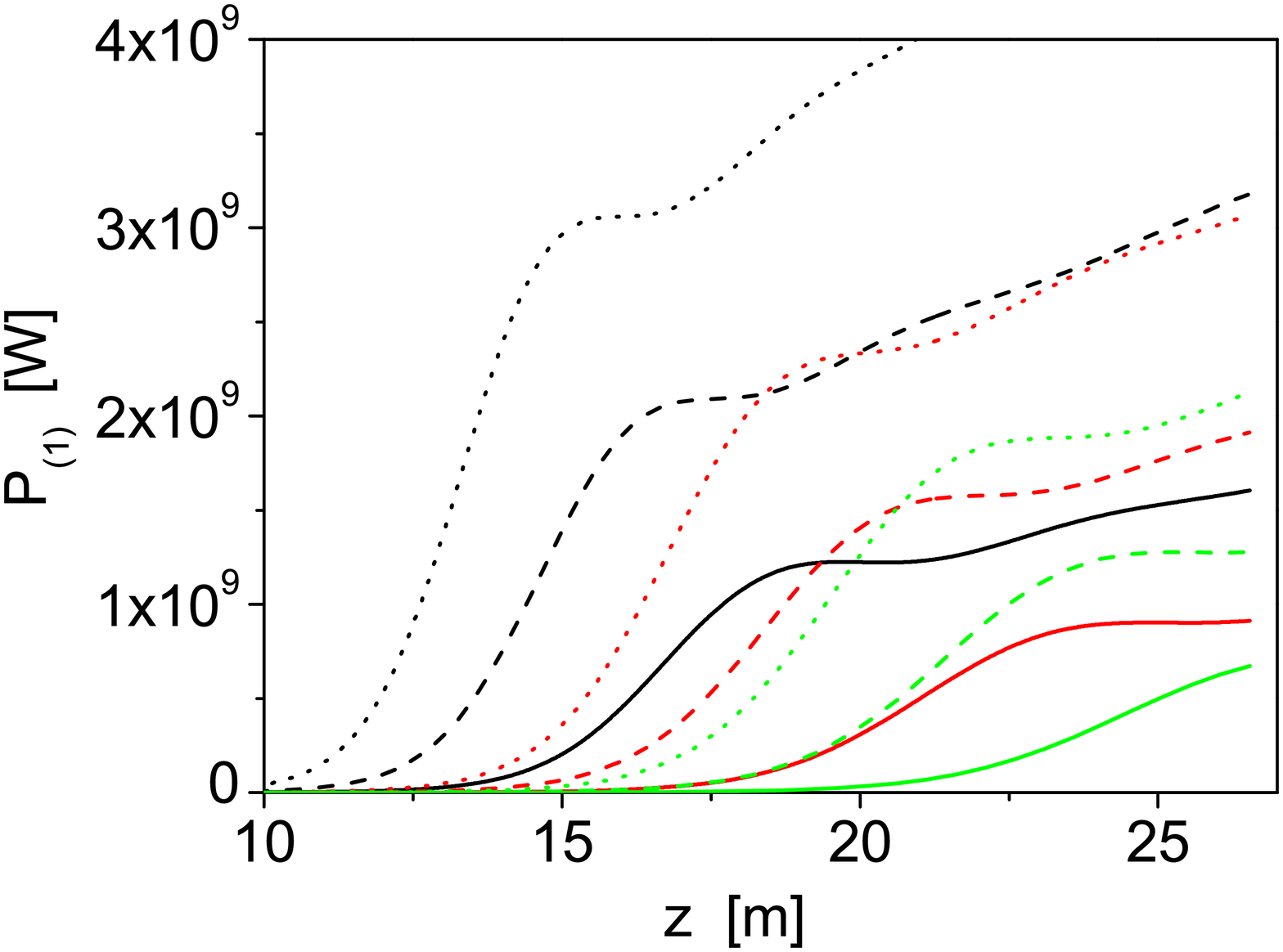}
\includegraphics[width=0.5\textwidth]{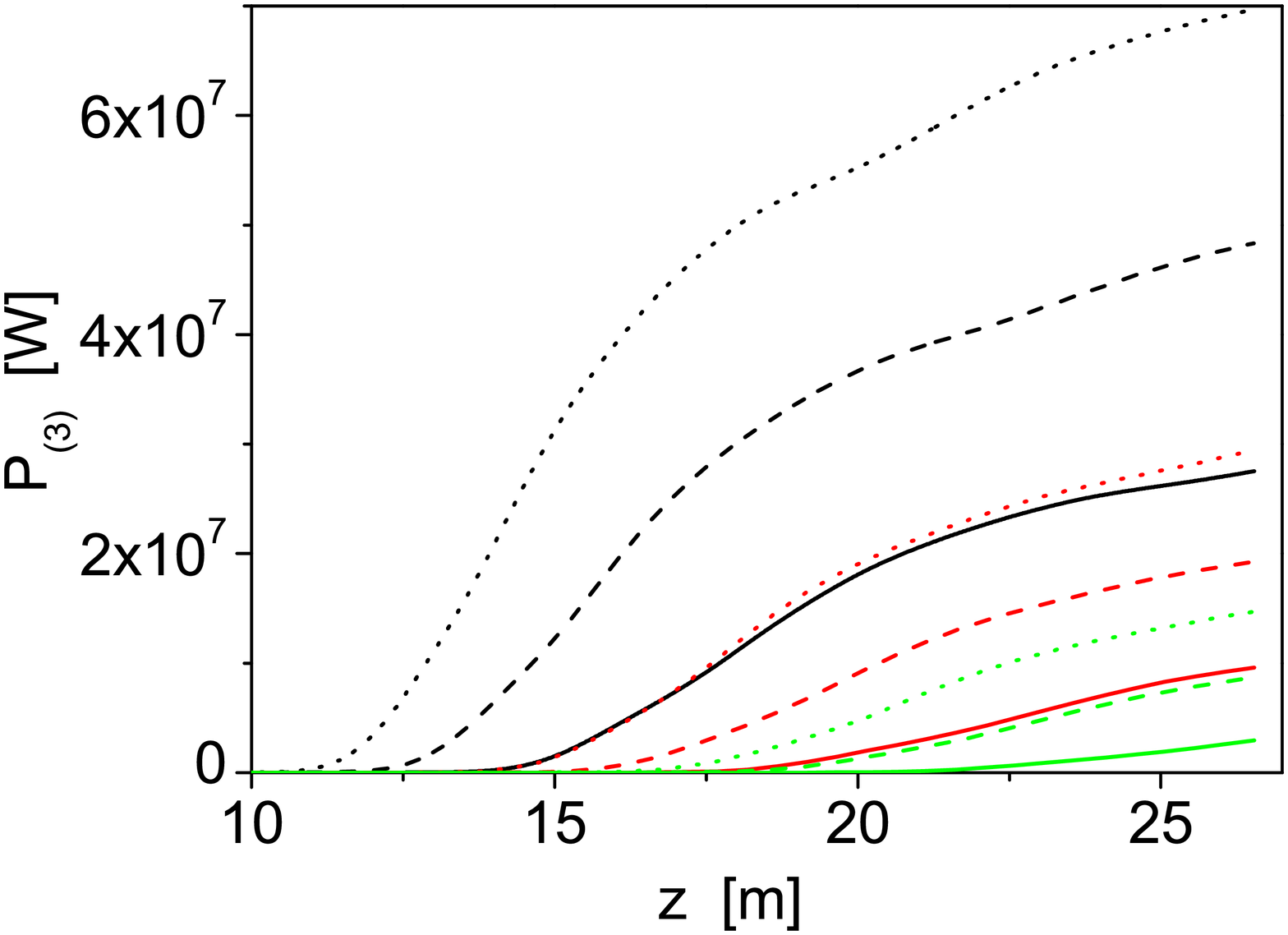}

\caption{Evolution of the radiation power along undulator
for the fundamental harmonic (left plot) and for the 3rd harmonic
(right plot).
Color codes (black, red and green) refer to different emittance
$\epsilon _n = 0.5$, 1, and 1.5 mm-mrad. Line styles
(solid, dash, and dot) refer to different values of peak current 1 kA, 1.5 kA,
and 2 kA.
Radiation wavelength is 8 nm.
Beta function is 10 m.
}
\label{fig:pz1}
\end{figure}

\begin{figure}[tb]

\includegraphics[width=0.5\textwidth]{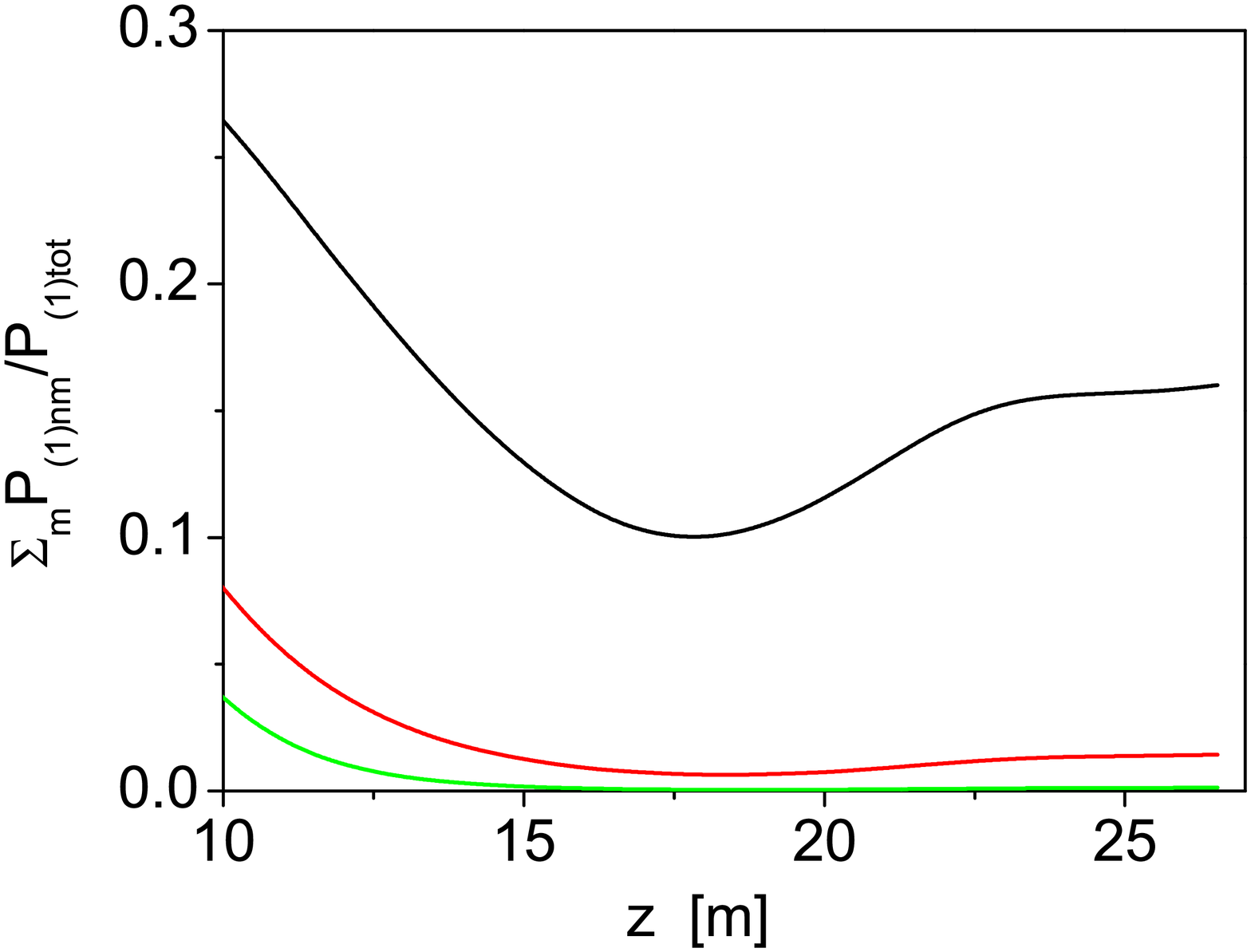}
\includegraphics[width=0.5\textwidth]{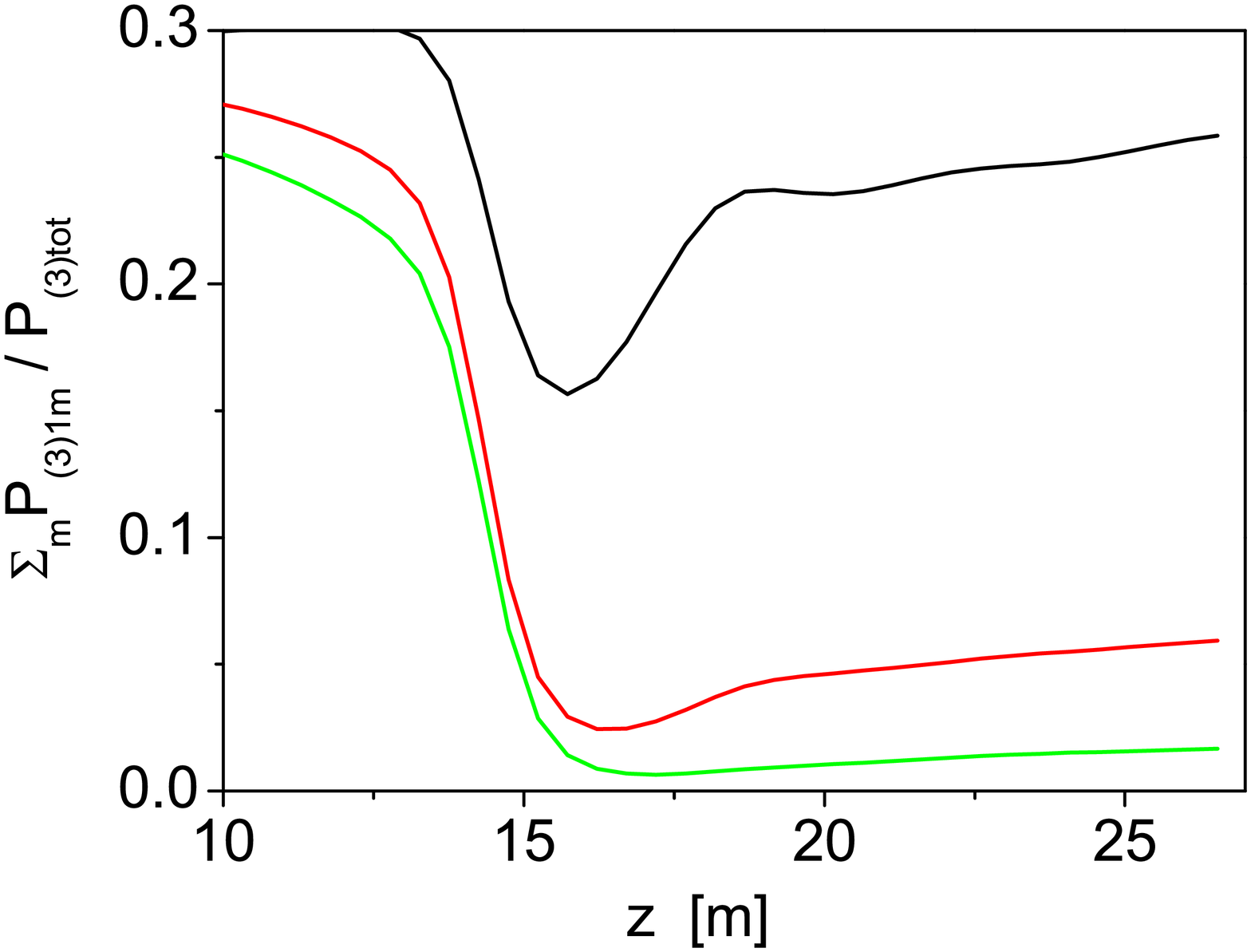}

\caption{
Partial contribution of the higher azimuthal modes for the fundamental harmonic
(left plot) and for the 3rd harmonic (right plot).
Black, red, and green curves refer to the modes
with $n = \pm 1$, $n = \pm 2$, and $n = \pm 3$, respectively.
Radiation wavelength is 8 nm.
Beta function is 10 m.
Beam current is 1.5 kA.
rms normalized emittance is 1 mm-mrad.
}

\label{fig:paz1toptot}

\end{figure}

\begin{figure}[tb]

\includegraphics[width=0.5\textwidth]{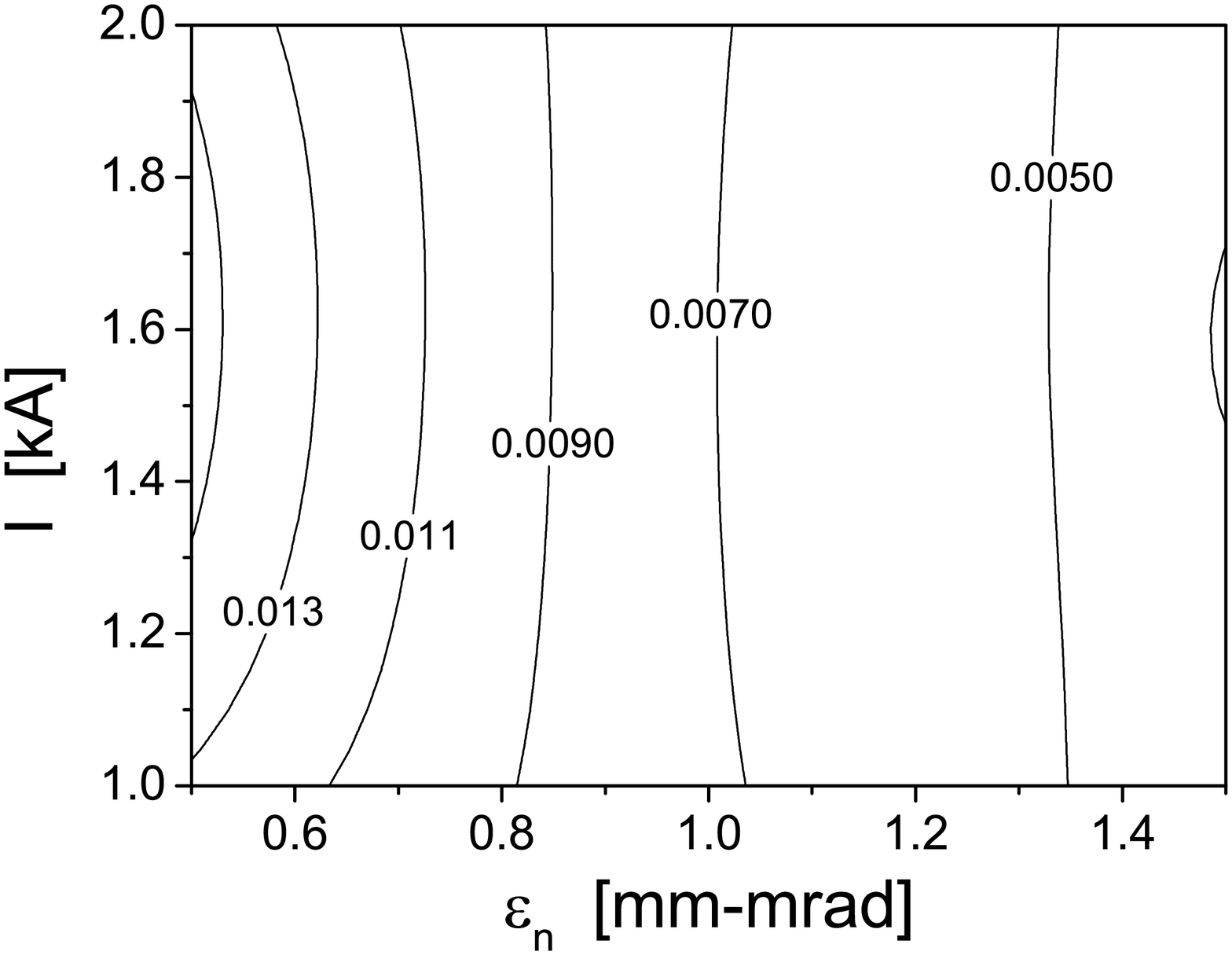}
\includegraphics[width=0.5\textwidth]{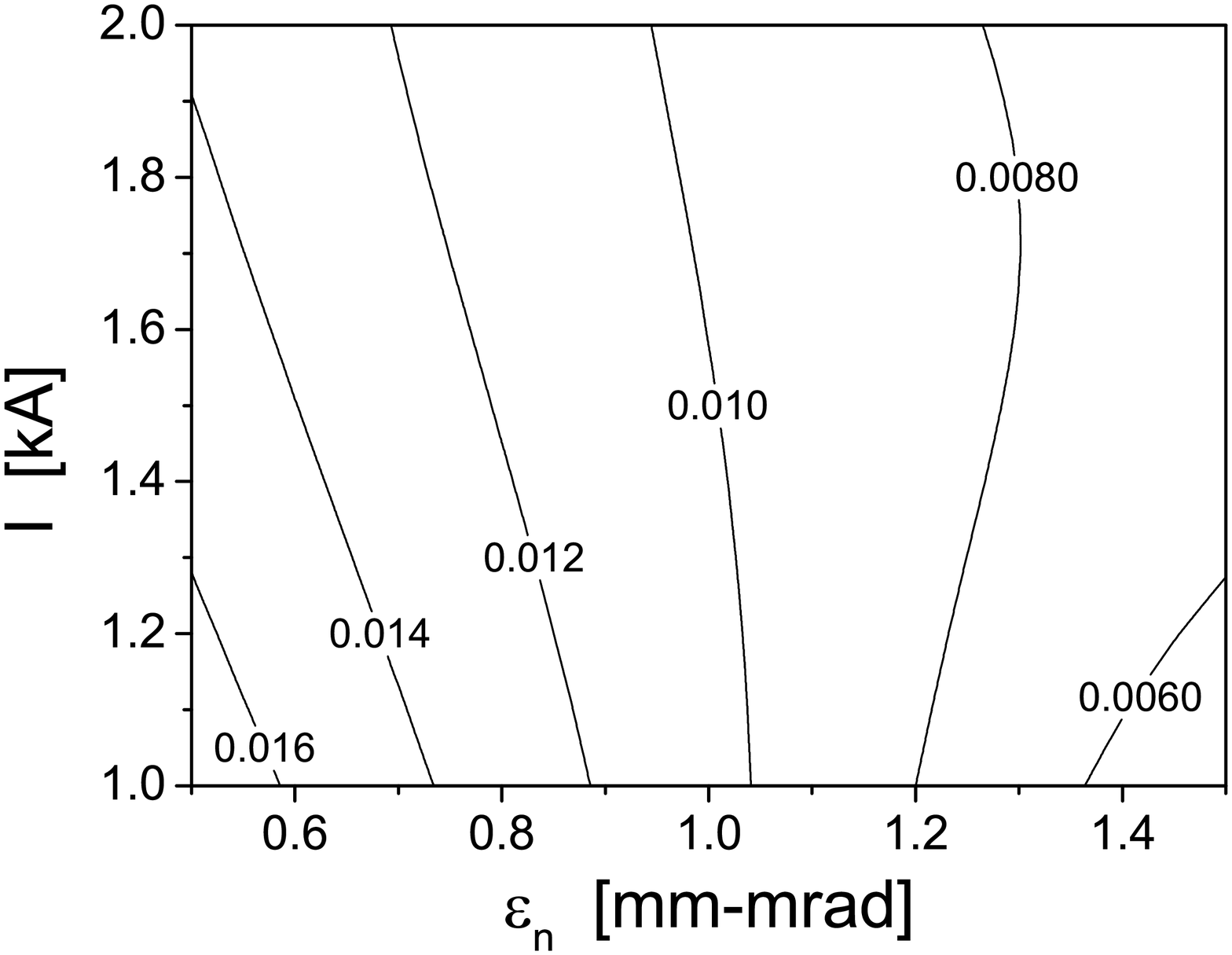}

\caption{
Partial contribution of the 3rd harmonic to the total power versus peak beam
current and emittance. Left and right plot refer to the saturation point and the
undulator end, respectively.
Radiation wavelength is 8 nm.
Beta function is 10 m.
Beam current is 1.5 kA.
rms normalized emittance is 1 mm-mrad.
}

\label{fig:partpow3}

\end{figure}

\begin{figure}[tb]

\includegraphics[width=0.5\textwidth]{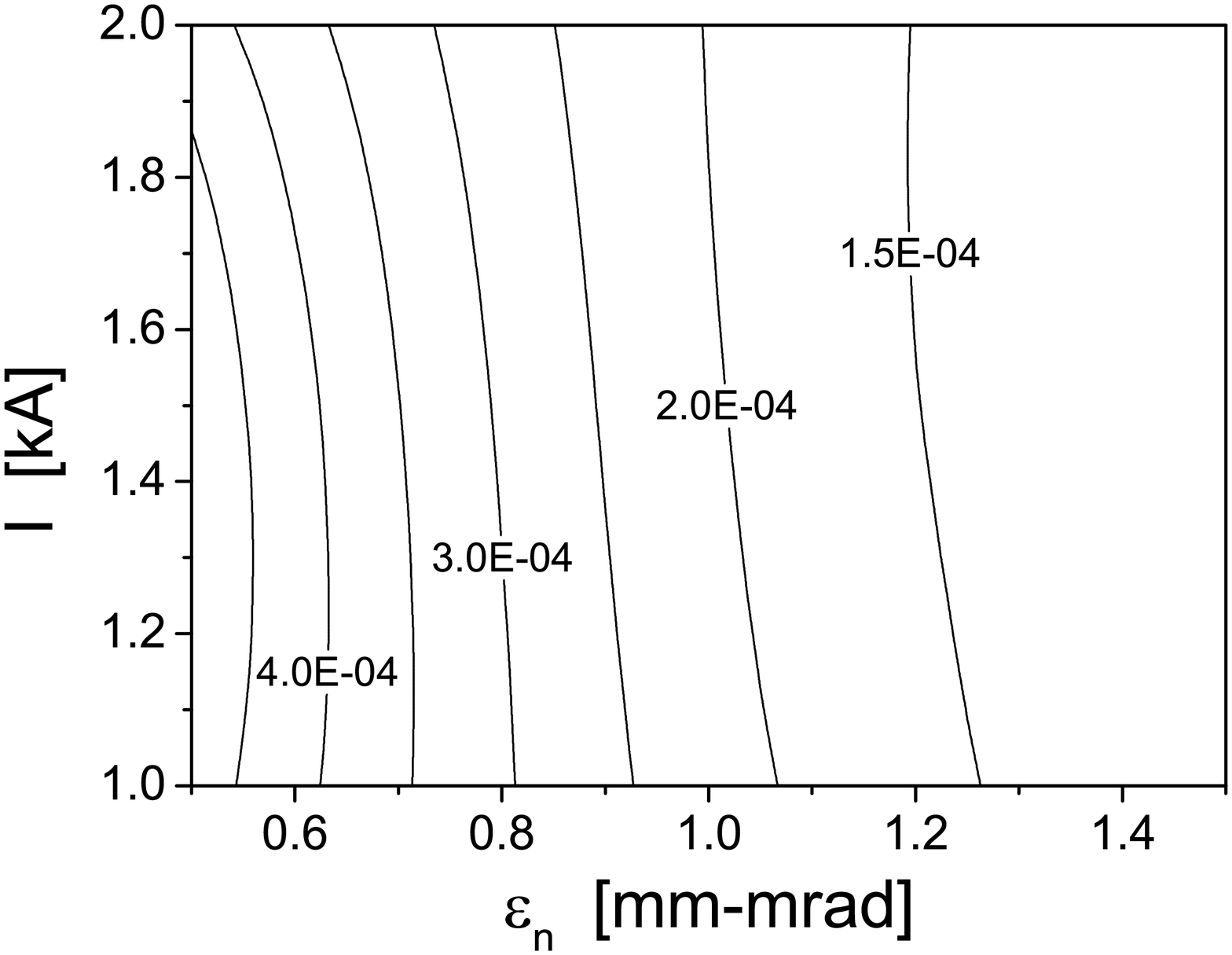}
\includegraphics[width=0.5\textwidth]{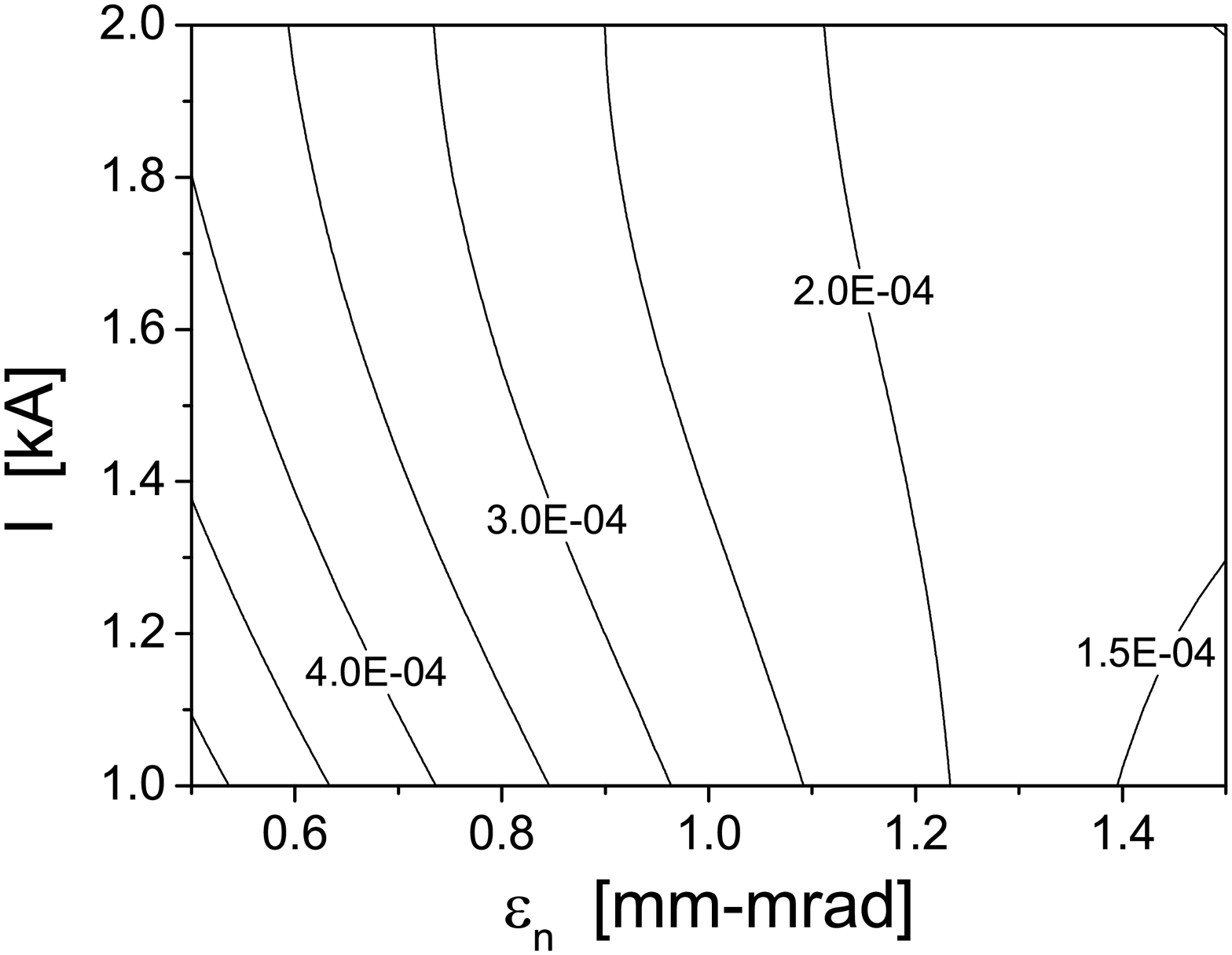}

\caption{
Partial contribution of the 5th harmonic to the total power versus peak beam
current and emittance. Left and right plot refer to the saturation point and the
undulator end, respectively.
Radiation wavelength is 8 nm.
Beta function is 10 m.
Beam current is 1.5 kA.
rms normalized emittance is 1 mm-mrad.
}

\label{fig:partpow5}

\end{figure}

\begin{figure}[tb]

\includegraphics[width=0.5\textwidth]{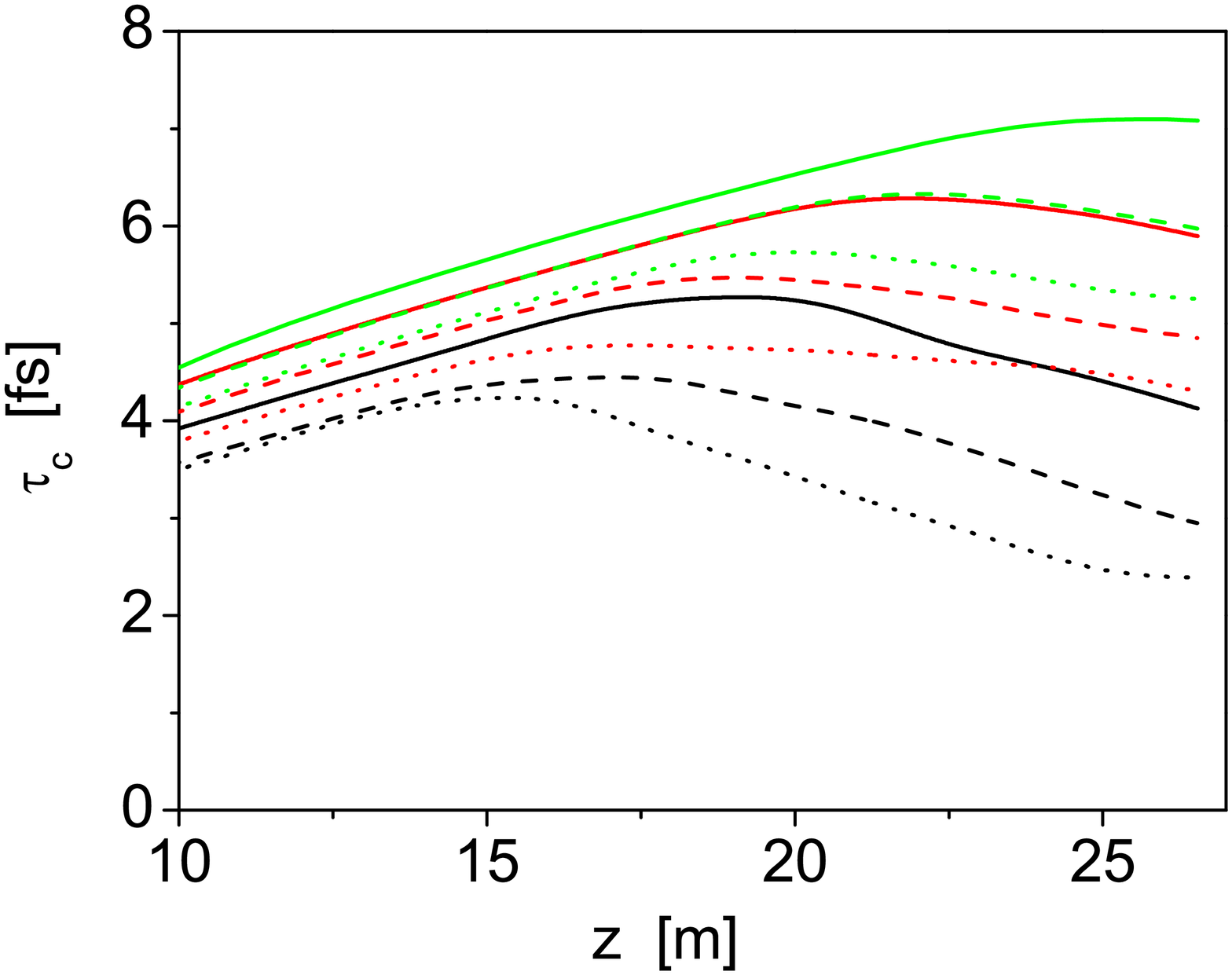}
\includegraphics[width=0.5\textwidth]{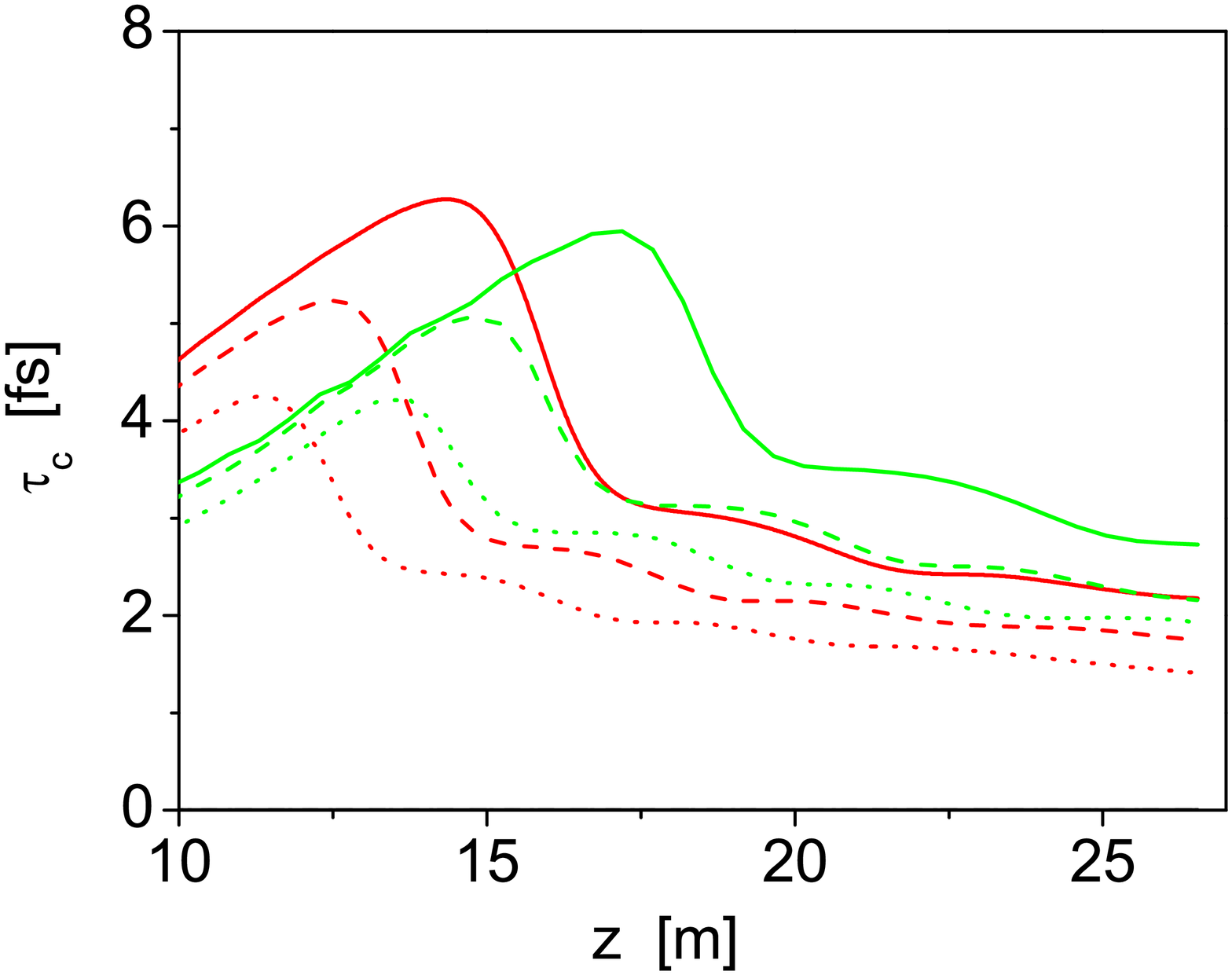}

\caption{Evolution along undulator of the coherence time of
the radiation at the fundamental harmonic (left plot) and at the
3rd harmonic (right plot).
Color codes (black, red and green) refer to different
emittance $\epsilon _n = 0.5$, 1, and 1.5 mm-mrad. Line styles (solid, dash,
and dot) refer to different values of peak current 1 kA, 1.5 kA, and 2 kA.
Radiation wavelength is 8 nm.
Beta function is 10 m.
}

\label{fig:tc1z}

\end{figure}

\begin{figure}[tb]

\includegraphics[width=0.5\textwidth]{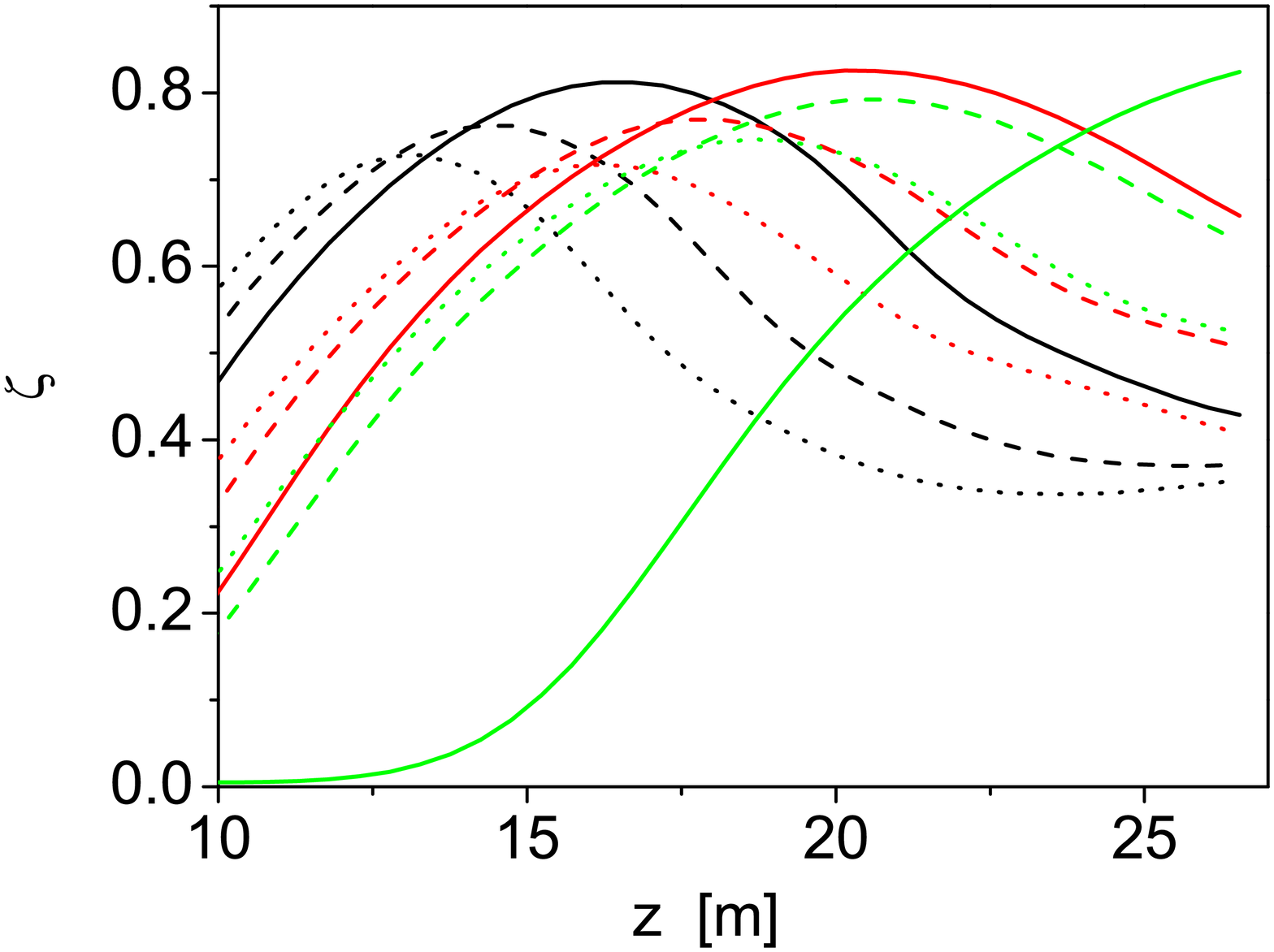}
\includegraphics[width=0.5\textwidth]{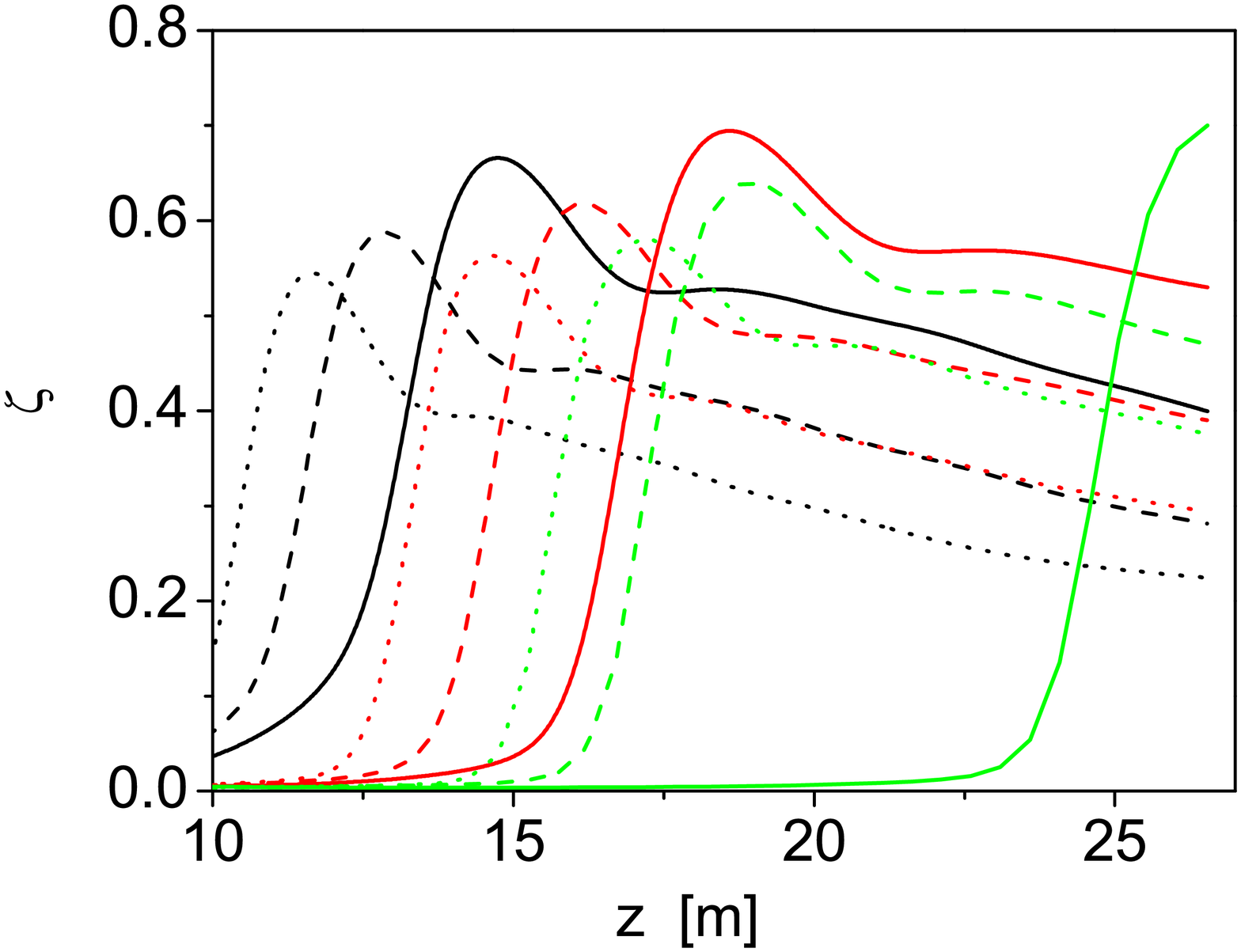}

\caption{Evolution along the undulator of the degree of transverse coherence of
the radiation. Left and right plots correspond to the fundamental
frequency (8 nm) and the 3rd harmonic (2.66 nm), respectively.
Color codes (black, red and green) refer to different emittance
$\epsilon _n = 0.5$, 1, and 1.5 mm-mrad. Line styles
(solid, dash, and dot) refer to different values of peak current 1 kA, 1.5 kA,
and 2 kA.
Radiation wavelength is 8 nm.
Beta function is 10 m.
}
\label{fig:dc1z}
\end{figure}

We present in Figure \ref{fig:pz1} evolution along the undulator of the
radiation power in the fundamental harmonic.
Higher values of the peak current and smaller emittances would allow to reach
higher radiation powers. When amplification process enters nonlinear stage, the
process of nonlinear harmonic generation takes place
\cite{harm-prst,hg-1,hg-2,hg-2a,hg-3,hg-4,hg-5,hg-6,kim-1,kim-2}.
Figures~\ref{fig:partpow3} and \ref{fig:partpow5} show relevant contribution to
the total power of the 3rd and the 5th harmonic.
The saturation point and the total undulator length (27 m) are chosen as
reference points. General observation is that the relative contribution of the
higher harmonic is higher for smaller values of the emittance. With the value
of the normalized emittance of 1 mm-mrad, partial contributions for the 3rd and
the 5th harmonic are in the range of $(0.7-1)\times 10^{-2}$ and $(2-2.5)\times
10^{-4}$, respectively. Note that this result is pretty much close to that
described by universal scaling with an appropriate correction for longitudinal
velocity spread derived in \cite{harm-prst}:

\begin{equation}
\frac{\langle W_3 \rangle}{\langle W_1 \rangle} \vert _{\mathrm{sat}}  =
0.094 \times \frac{K_3^2}{K_1^2} \ , \qquad
\frac{\langle W_5 \rangle}{\langle W_1 \rangle} \vert _{\mathrm{sat}} =
0.03 \times \frac{K_5^2}{K_1^2} \ .
\label{eq:sat35}
\end{equation}

\noindent Here $K_h = K(-1)^{(h-1)/2} [J_{(h-1)/2}(Q) - J_{(h+1)/2}(Q)]$, $Q =
K^2/[2(1+K^2)]$, $K$ is rms undulator parameter, and $h$ is an odd integer -
harmonic number.

\subsection{Temporal coherence}

In the framework of the one-dimensional model the coherence time at the
saturation point is described in terms of FEL parameter $\rho _{1D}$
\cite{boni-rho} and a number of cooperating electrons $N_{\mathrm{c}} = I/(e\rho
_{1D}\omega )$ \cite{book}:

\begin{displaymath}
\tau _{\mathrm{c}} \simeq
\frac{1}{\rho _{1D}\omega }
\sqrt{ \frac{\pi \ln N_{\mathrm{c}} }{18} } \ .
\end{displaymath}

The coherence time of higher harmonics at the saturation point and
in the post-saturation amplification stage scales inversely proportional to the
harmonic number, while relative spectrum bandwidth remains constant with the
harmonic number.

Blue curve in Fig.~\ref{fig:dctpz1510} shows evolution of the coherence time
which is typical for all SASE FELs. In the high gain linear regime it increases
as a square root of undulator length. It reaches maximum value just before
saturation point, and then it drops down. Figures~\ref{fig:tc1z} show the coherence
time for the whole parameter range for the fundamental harmonic. Coherence time
for the higher harmonics can be derived using scaling described above.

\subsection{Spatial coherence}

Figure~\ref{fig:dc1z} presents an overview of the degree of transverse
coherence in the considered parameter space.
In our studies of coherent properties of FELs \cite{coherence-oc} we have found that for an
optimized SASE FEL the degree of transverse coherence can be as high as 0.96.
One can see from Fig.~\ref{fig:dc1z} that in the considered cases the degree of transverse coherence
for the 1st harmonic is visibly lower.

We should distinguish two effects limiting the degree of transverse coherence at FLASH.
The first one is called mode degeneration and was intensively discussed in this paper. This
physical phenomena takes place at large values of the diffraction
parameter \cite{book}. Figure\ref{fig:paz1toptot} shows contribution of higher
azimuthal modes to the total power for specific example of emittance 1 mm-mrad
and peak current 1.5 kA . Contribution of the first
azimuthal modes falls down in the high gain linear regime, but to the value of
12\% only, and then starts to grow in the nonlinear regime, and reaches the
value of 16\% at the undulator end.

The second effect is connected with a finite longitudinal coherence, it was
discovered in \cite{trcoh-oc} and discussed in
\cite{coherence-oc,coherence-njp}. The essence of the effect is a superposition
of mutually incoherent fields produced by different longitudinally uncorrelated
parts of the electron  bunch. In the exponential gain regime this effect is
relatively weak, but it prevents a SASE FEL from reaching full transverse
coherence even in the case when only one transverse eigenmode survives
\cite{trcoh-oc}. In the deep nonlinear regime beyond FEL saturation, this
effect can be strong and can lead to a significant degradation of the degree of
transverse coherence \cite{coherence-oc,coherence-njp}. In particular, as one
can see from Fig.~\ref{fig:dc1z}, this effect limits the degree of transverse
coherence to the value about 50\% when FLASH operates in the deep nonlinear
regime.

Higher harmonics are derived from the nonlinear process governed by the
fundamental harmonic. As a result, coherence properties of the harmonics
follow the same tendencies as the fundamental, but with visibly lower degree of
transverse coherence \cite{coh-fel2012}.

Note that an easiest way to dramatically improve the transverse coherence would be to decrease the beam current
such that the saturation is achieved at the very end of the undulator. This would eliminate not only the degradation
in the deep nonlinear regime, but would also improve the mode selection process because the diffraction parameter
is then reduced while the velocity spread due to emittance is increased. According to our expectations,
the degree of transverse coherence might reach the value around 90\% in this regime. Such a regime was
realized at FLASH on user`s demand, but it is not typical for the machine operation
because the peak power is low due to a low peak current.

One can also suppress the unwanted effects in the deep nonlinear regime by kicking the electron beam at the
saturation point (or, close to it) when the peak current is high.
Then one can still have a high power and an improved (about 70-80\%) degree
of transverse coherence. Further improvement could be achieved by reducing beta-function (thus improving the mode
selection as discussed above) but this would
cause some technical problems that we do not discuss in this paper.

\subsection{Pointing stability and mode degeneration}

\begin{figure}[tb]

\includegraphics[width=0.5\textwidth]{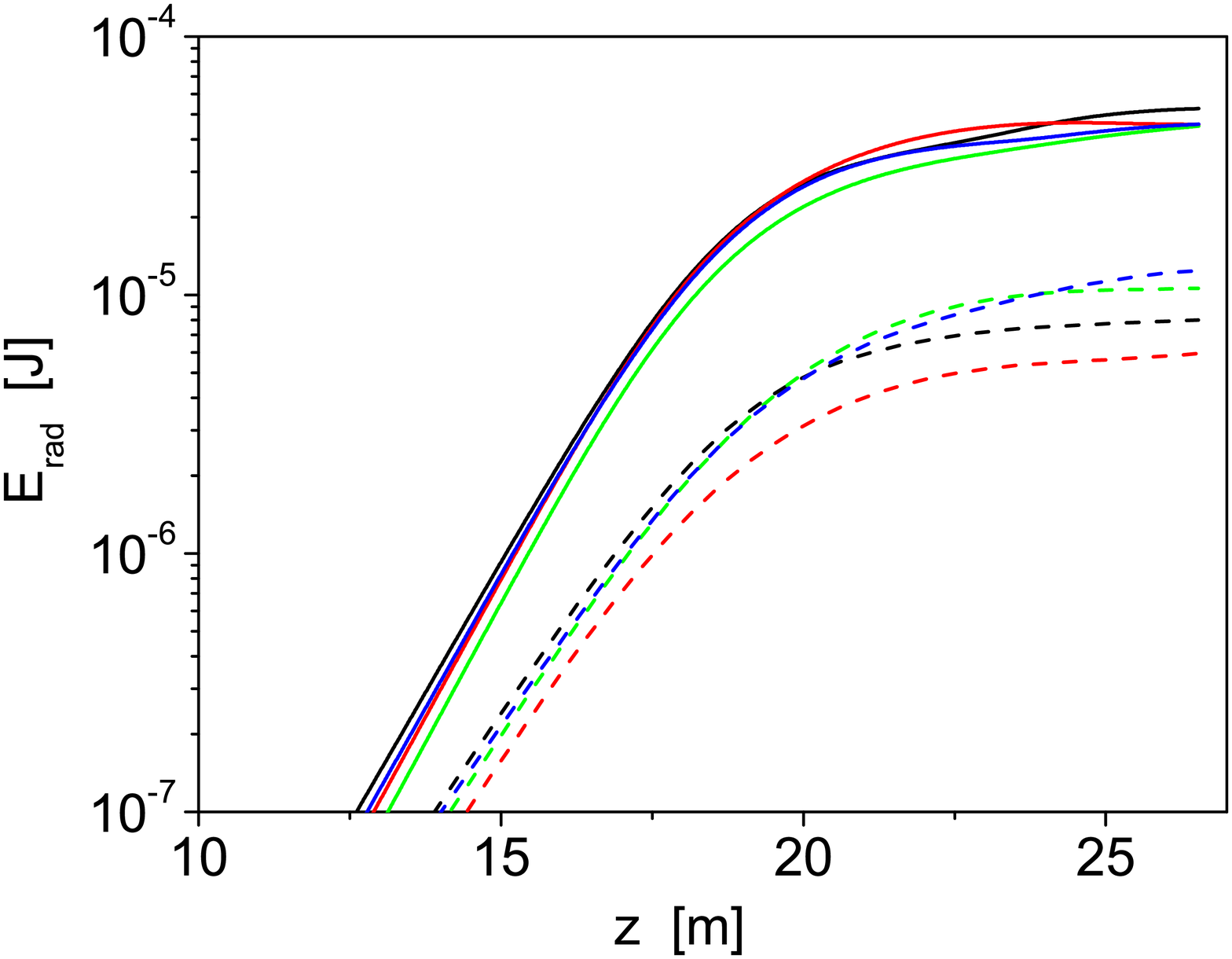}
\includegraphics[width=0.5\textwidth]{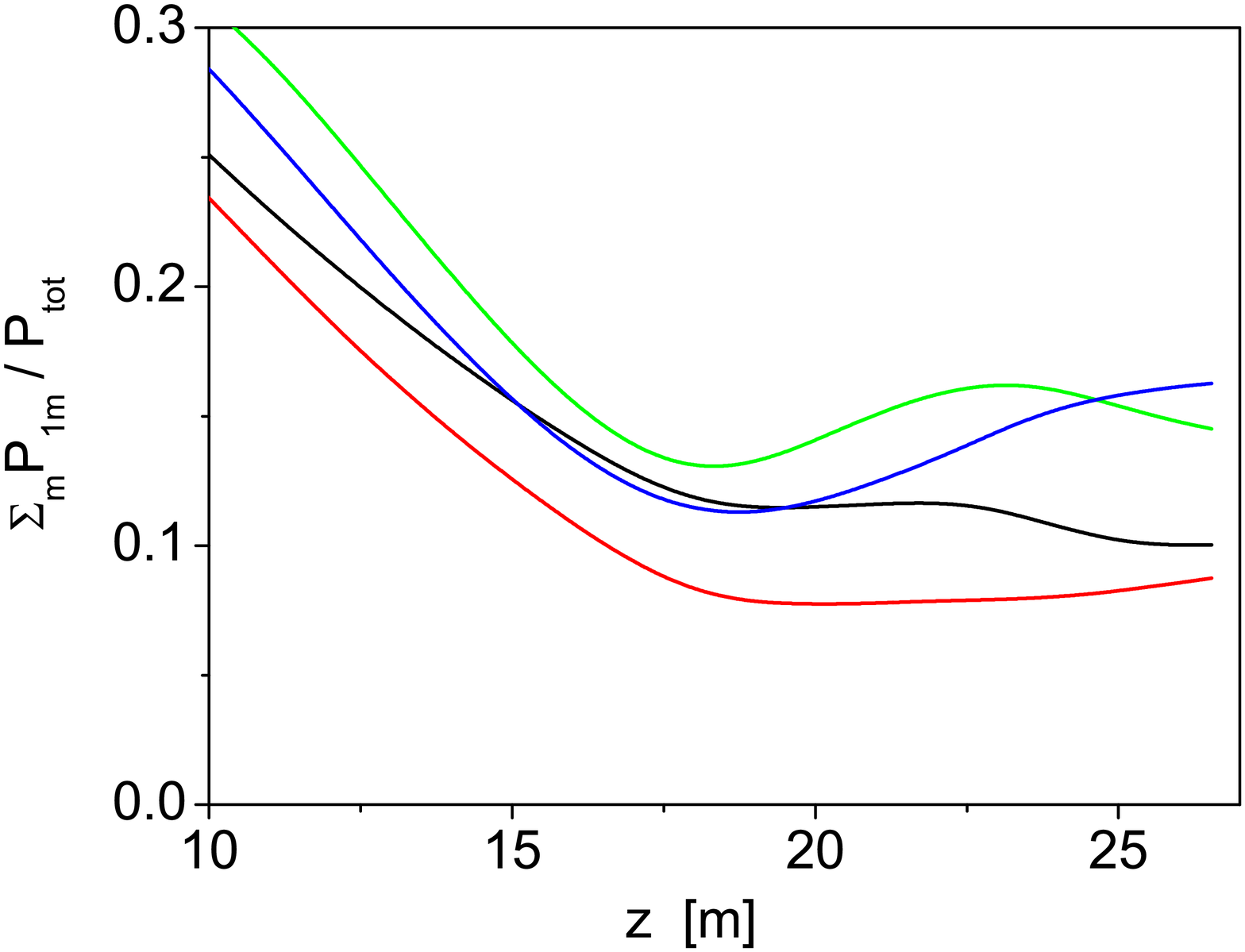}

\caption{
Left plot: evolution of the energy in the radiation pulse versus undulator length.
Color codes (black to blue) correspond to different shots. Line style correspond
to the total energy in the azimuthally symmetric $\sum TEM_{0m}$ modes (solid lines), and
in of the first azimuthal $\sum  TEM_{1m}$ (dashed lines).
Right plot:
partial contribution of the first azimuthal modes
to the total radiation power, $\sum P_{1m}/P_{tot}$.
Radiation wavelength is 8 nm.
Beta function is 10 m.
Beam current is 1.5 kA.
rms normalized emittance is 1 mm-mrad.
}

\label{fig:pz-60uj}

\end{figure}

\begin{figure}[tb]

\includegraphics[width=0.5\textwidth]{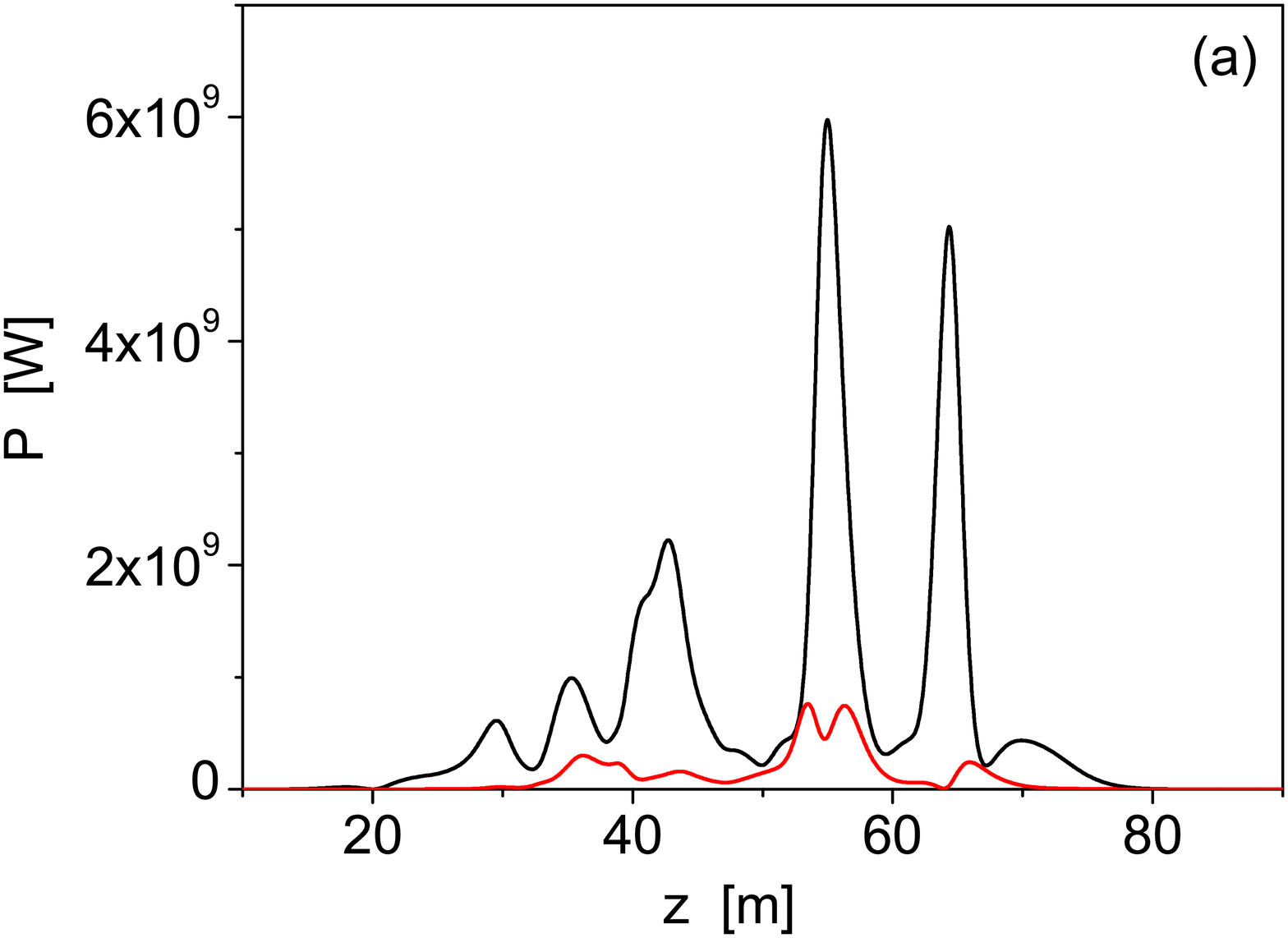}
\includegraphics[width=0.5\textwidth]{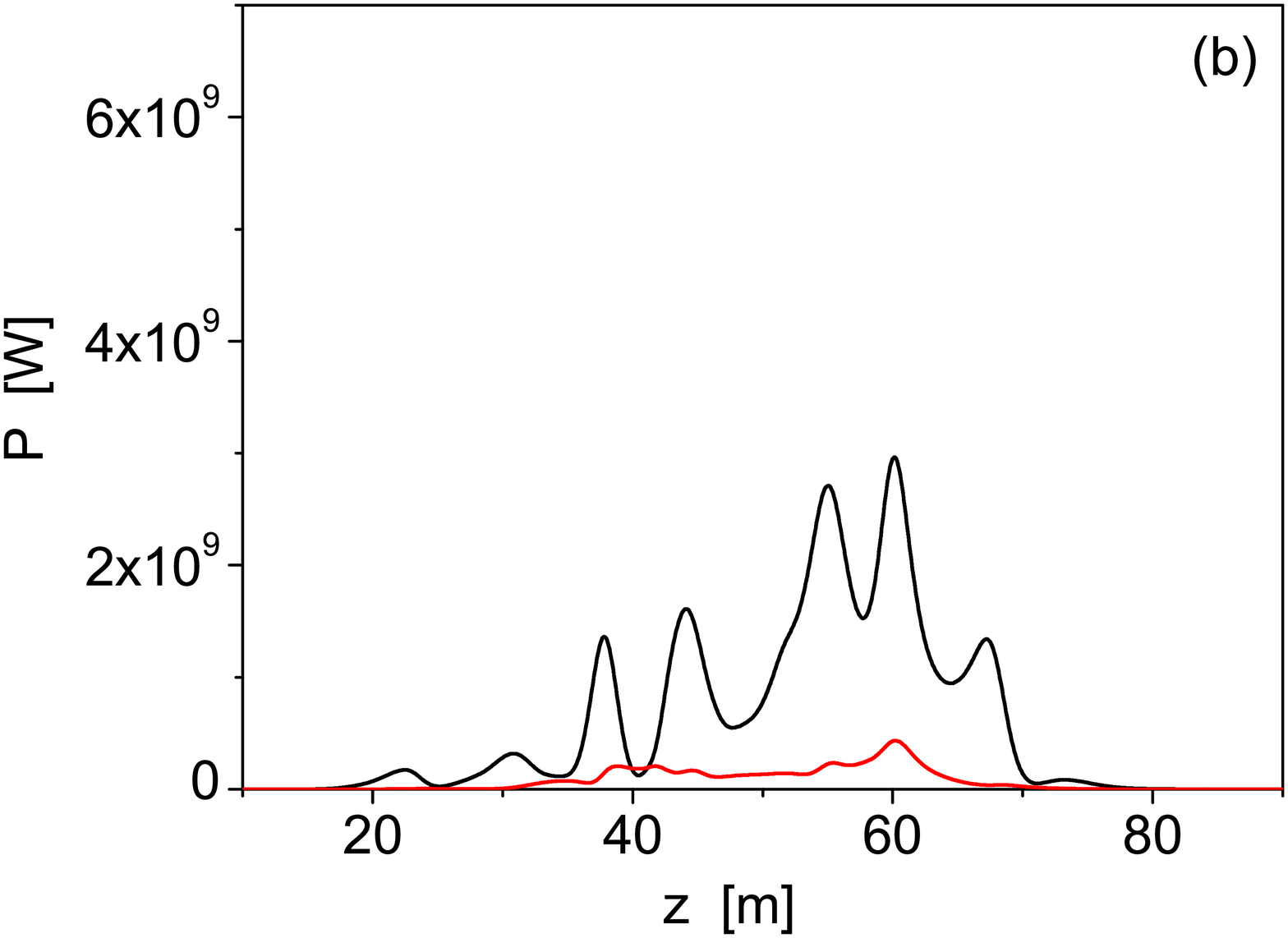}

\includegraphics[width=0.5\textwidth]{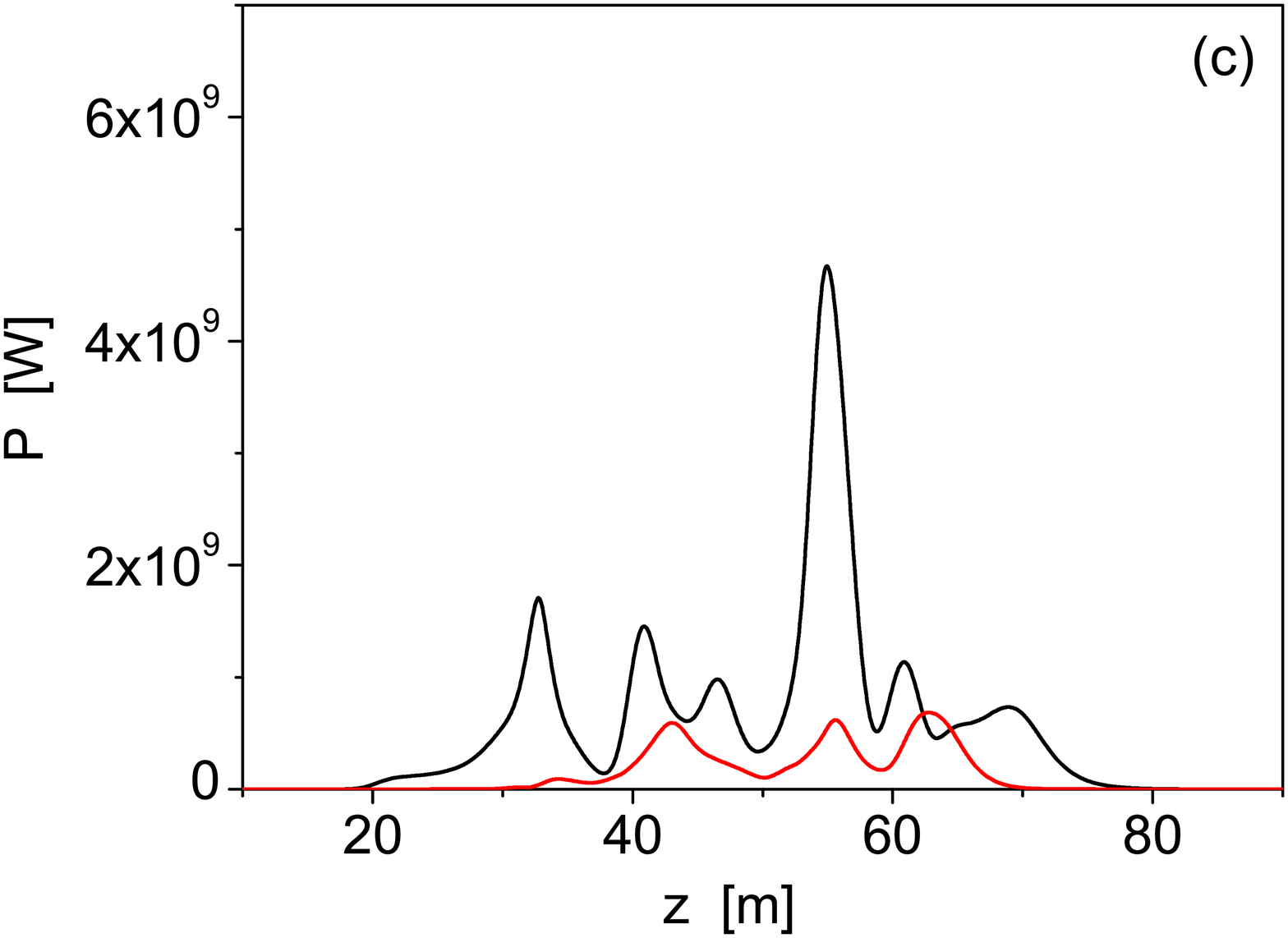}
\includegraphics[width=0.5\textwidth]{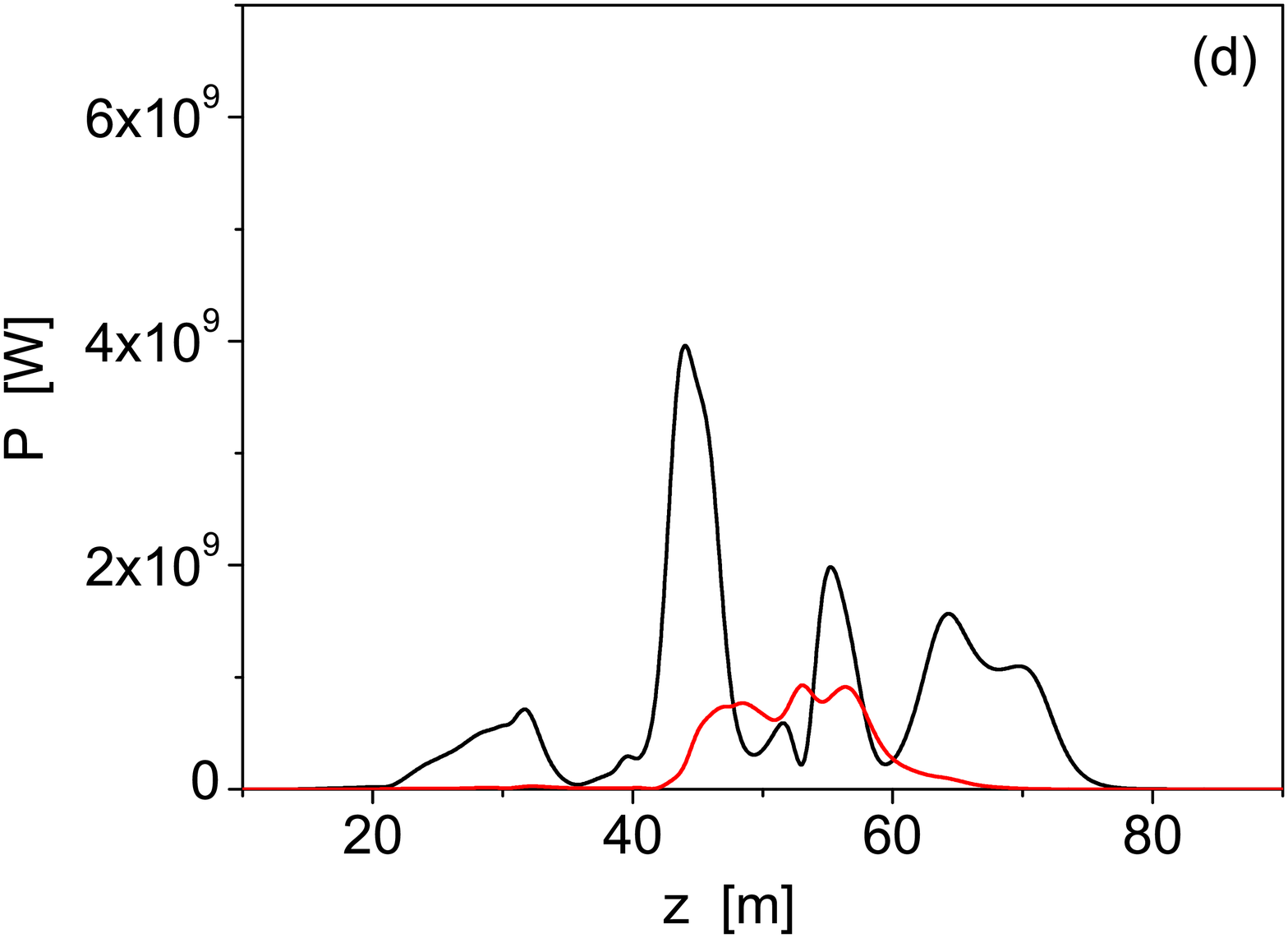}

\caption{
Temporal structure of four radiation pulses. Black lines show the power of the
azimuthally symmetric modes, and the curve in the red color show the
power of the first azimuthal modes.
Radiation
wavelength is 8 nm. Beta function is 10 m. Beam current is 1.5 kA. rms
normalized emittance is 1 mm-mrad. Undulator length is 27 m. }

\label{fig:temporal-60uj}

\end{figure}

\begin{figure}[tb]

\includegraphics[width=0.18\textwidth]{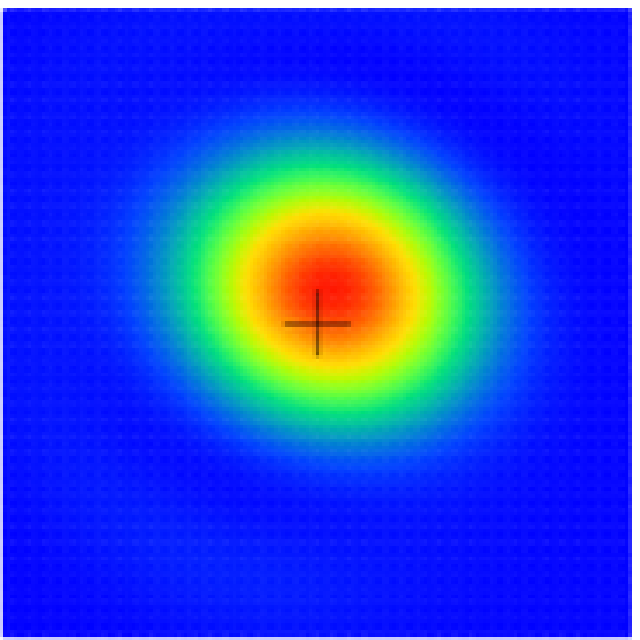}
\includegraphics[width=0.18\textwidth]{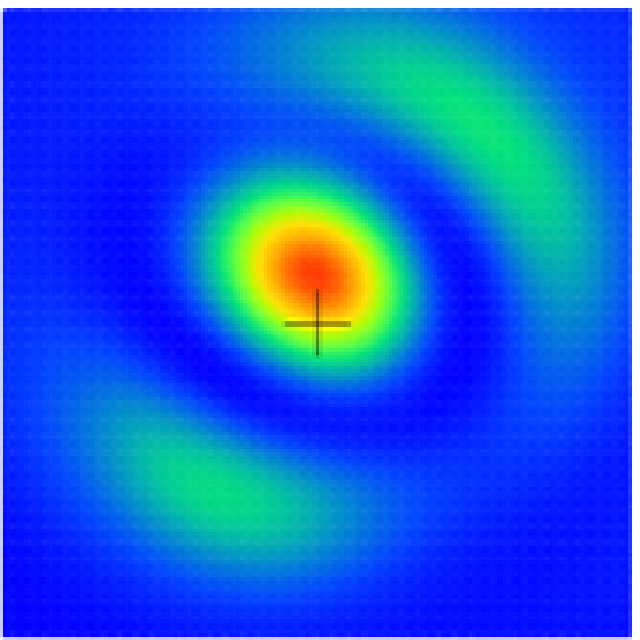}
\includegraphics[width=0.18\textwidth]{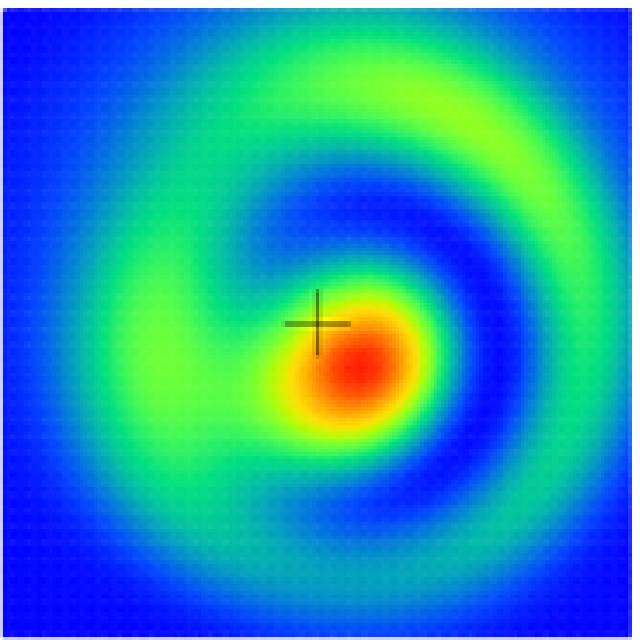}
\includegraphics[width=0.18\textwidth]{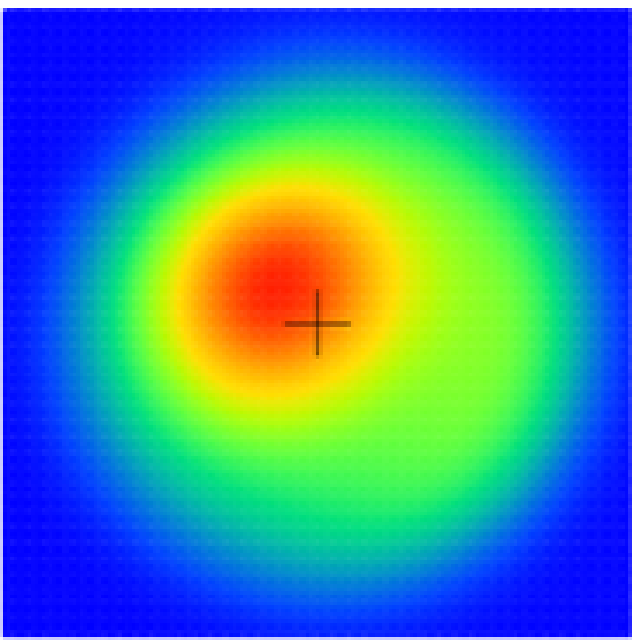}
\includegraphics[width=0.18\textwidth]{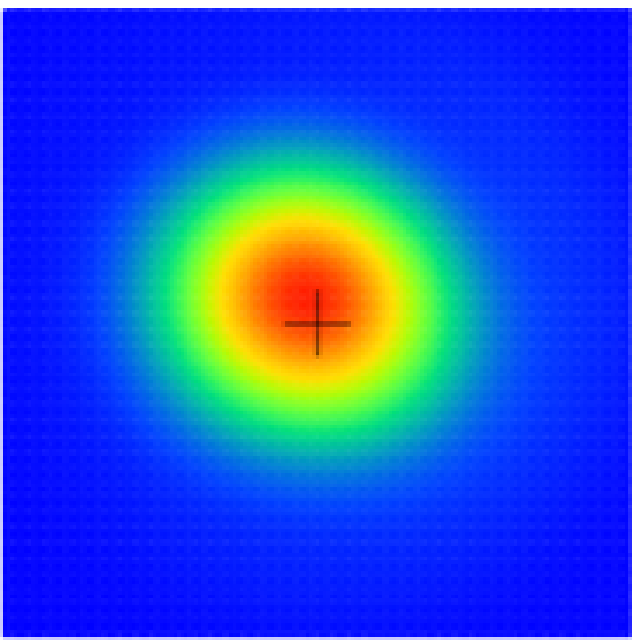}

(a)

\includegraphics[width=0.18\textwidth]{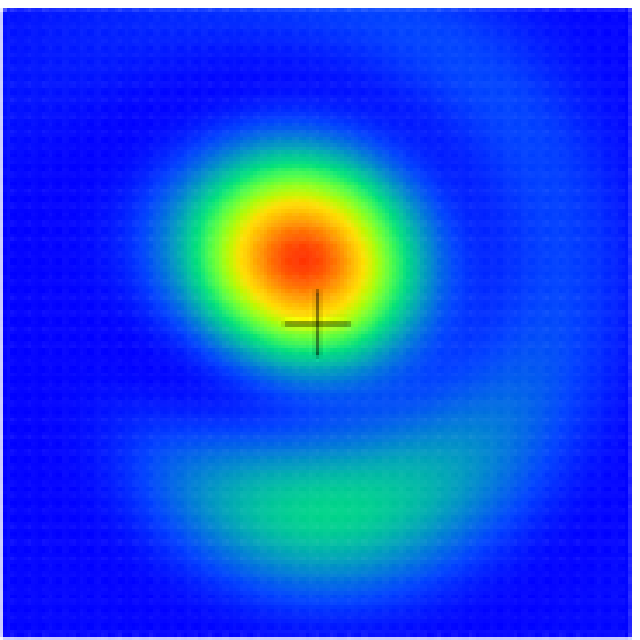}
\includegraphics[width=0.18\textwidth]{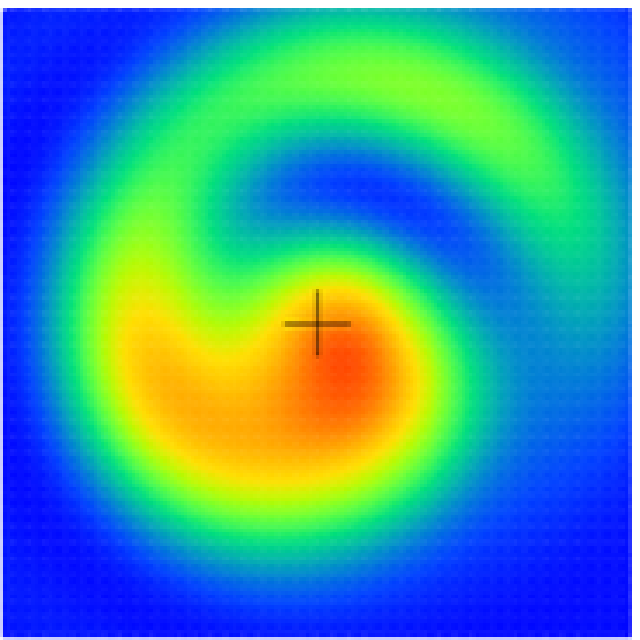}
\includegraphics[width=0.18\textwidth]{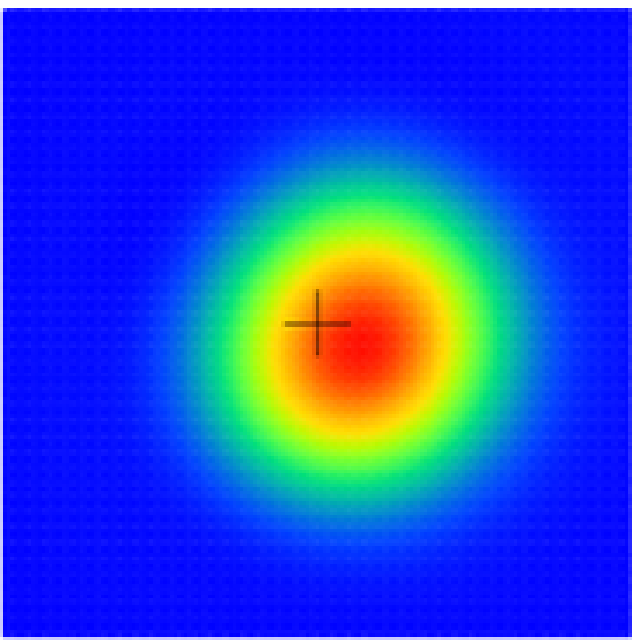}
\includegraphics[width=0.18\textwidth]{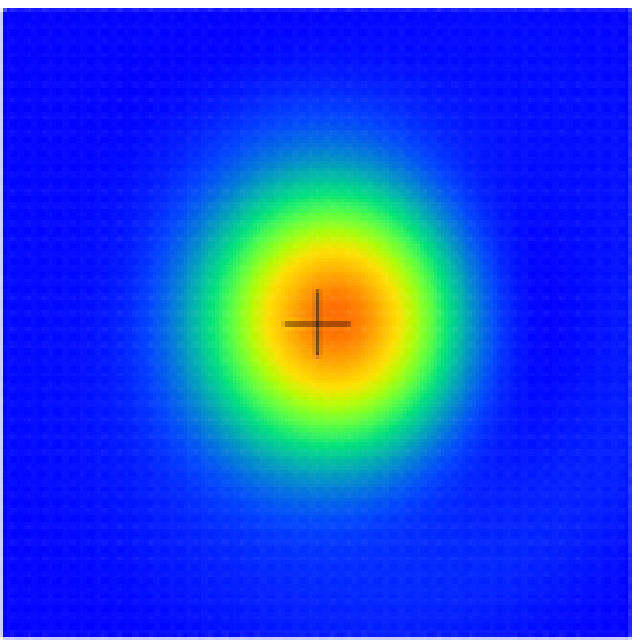}
\includegraphics[width=0.18\textwidth]{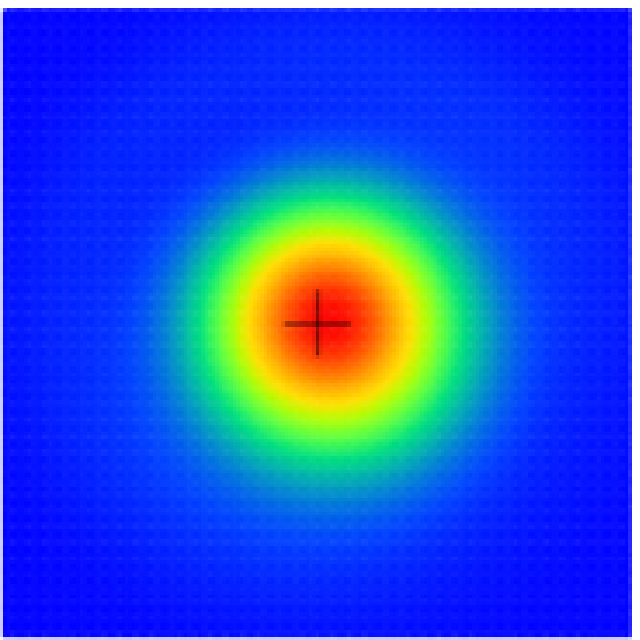}

(b)

\includegraphics[width=0.18\textwidth]{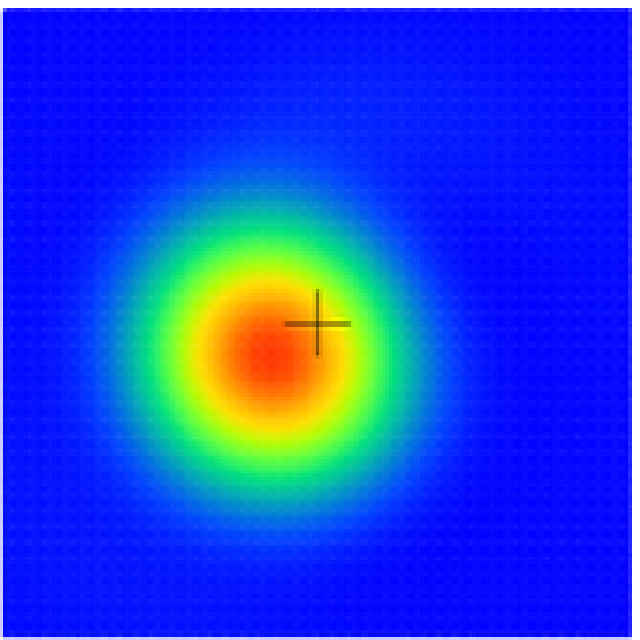}
\includegraphics[width=0.18\textwidth]{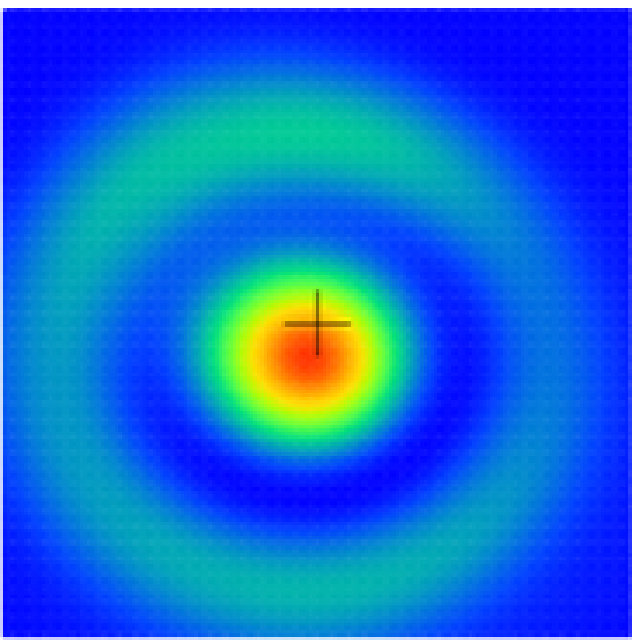}
\includegraphics[width=0.18\textwidth]{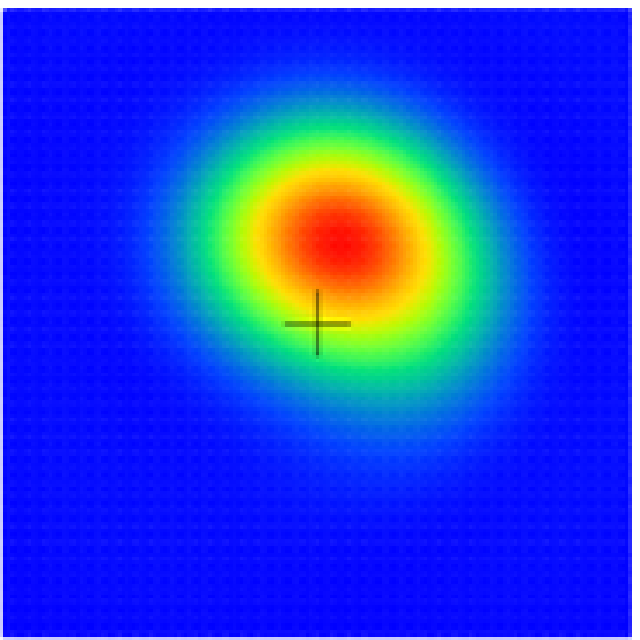}
\includegraphics[width=0.18\textwidth]{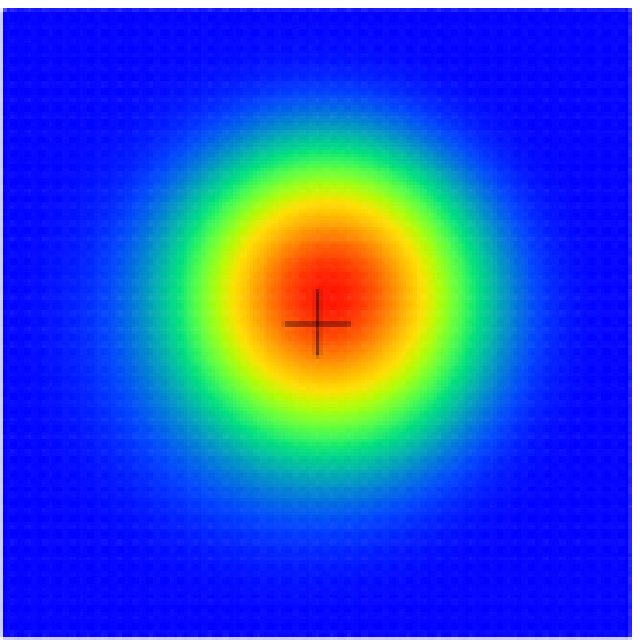}
\includegraphics[width=0.18\textwidth]{s06.eps}

(c)

\includegraphics[width=0.18\textwidth]{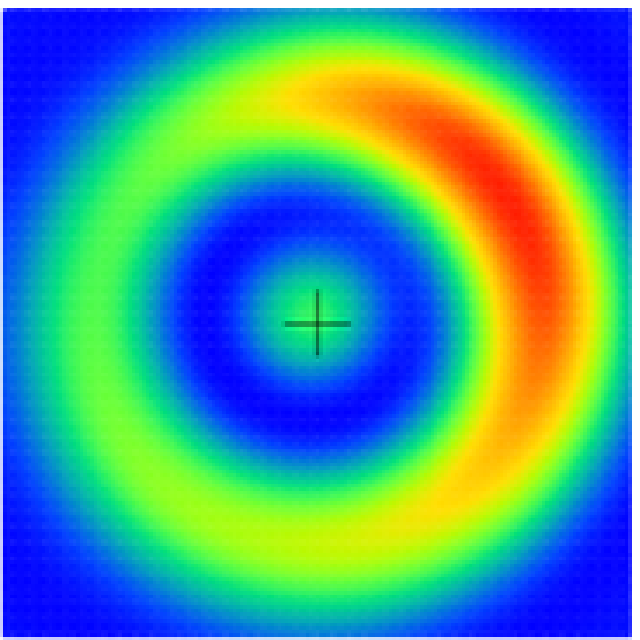}
\includegraphics[width=0.18\textwidth]{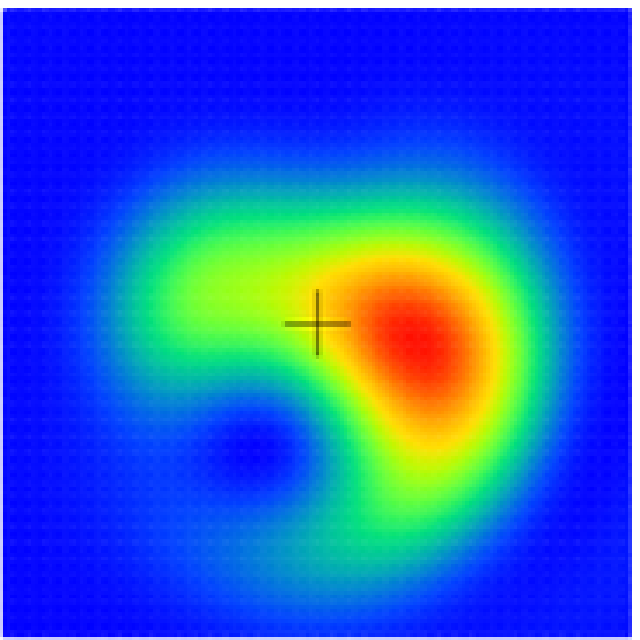}
\includegraphics[width=0.18\textwidth]{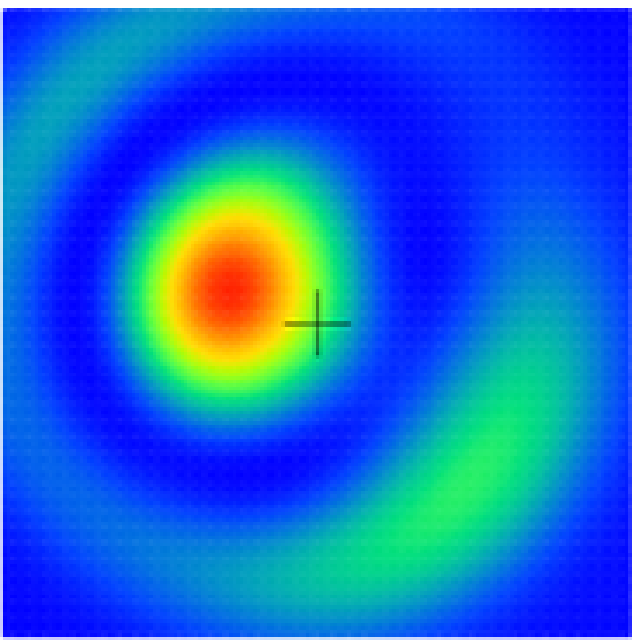}
\includegraphics[width=0.18\textwidth]{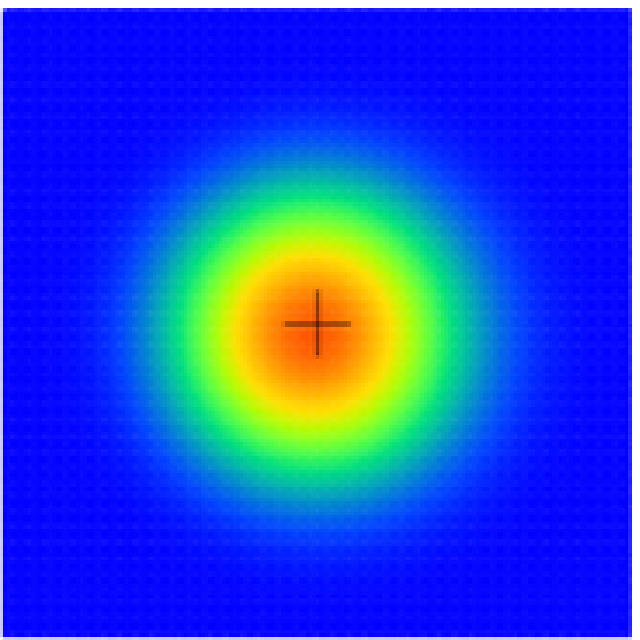}
\includegraphics[width=0.18\textwidth]{s06.eps}

(d)

\caption{
Profiles of the radiation intensity in the far zone.
Rows (a to d) correspond to specific shots with temporal structure
presented in Fig.~\ref{fig:temporal-60uj} (plots a to d). Profiles on the
right-hand side show average intensity over full pulse. Profiles 1 to 4 from the
left-hand size show intensity distribution of selected slices corresponding to
the time 40 fs, 50 fs, 60 fs, and 70 fs, respectively. Cross denotes
geometrical center of the radiation intensity averaged over many shots.
Radiation wavelength is 8 nm. Beta function is 10 m. Beam current is 1.5 kA.
rms normalized emittance is 1 mm-mrad. Undulator length is 27 m. }

\label{fig:profiles-60uj}

\end{figure}

\begin{figure}[tb]

\includegraphics[width=0.5\textwidth]{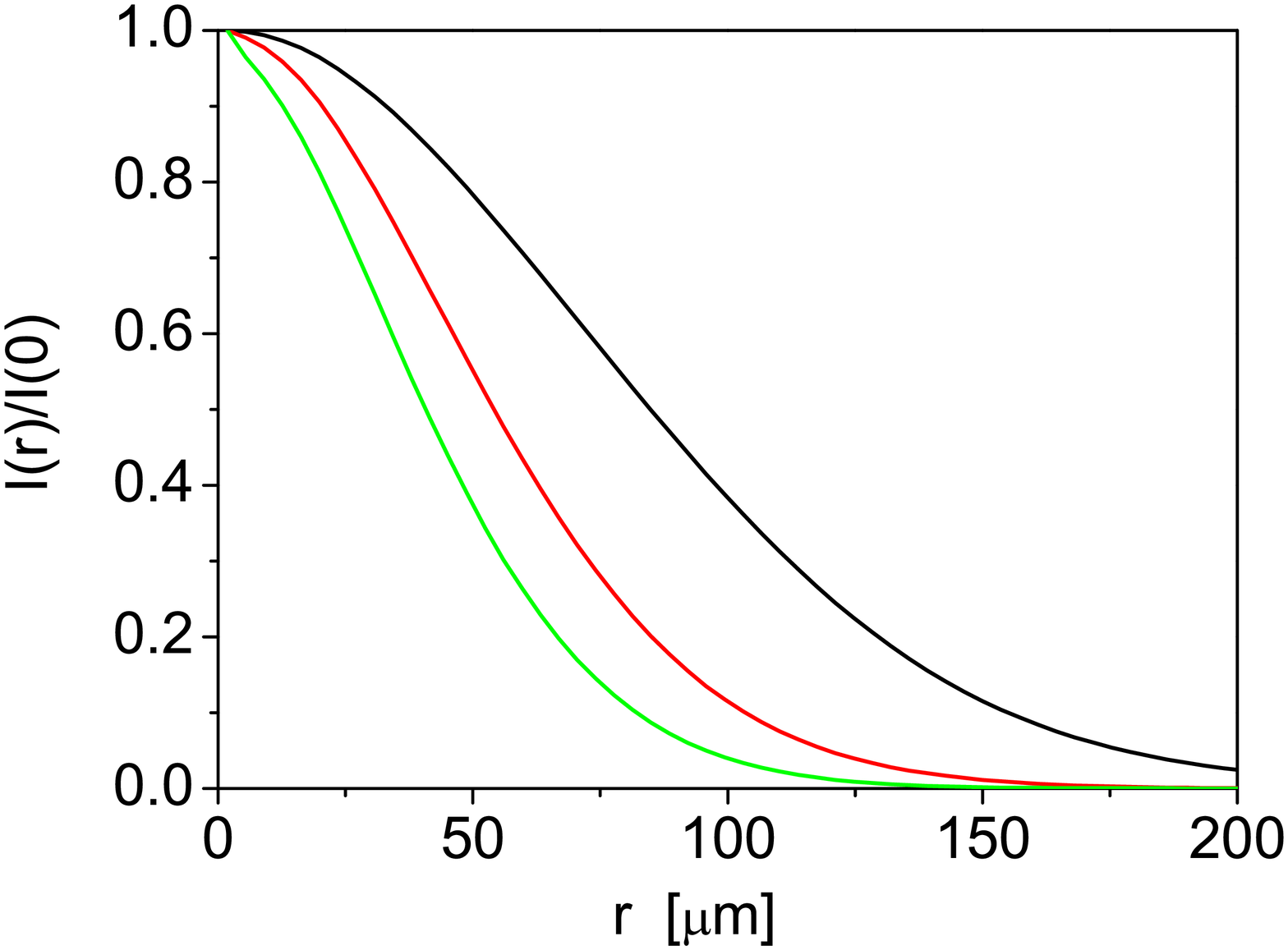}
\includegraphics[width=0.5\textwidth]{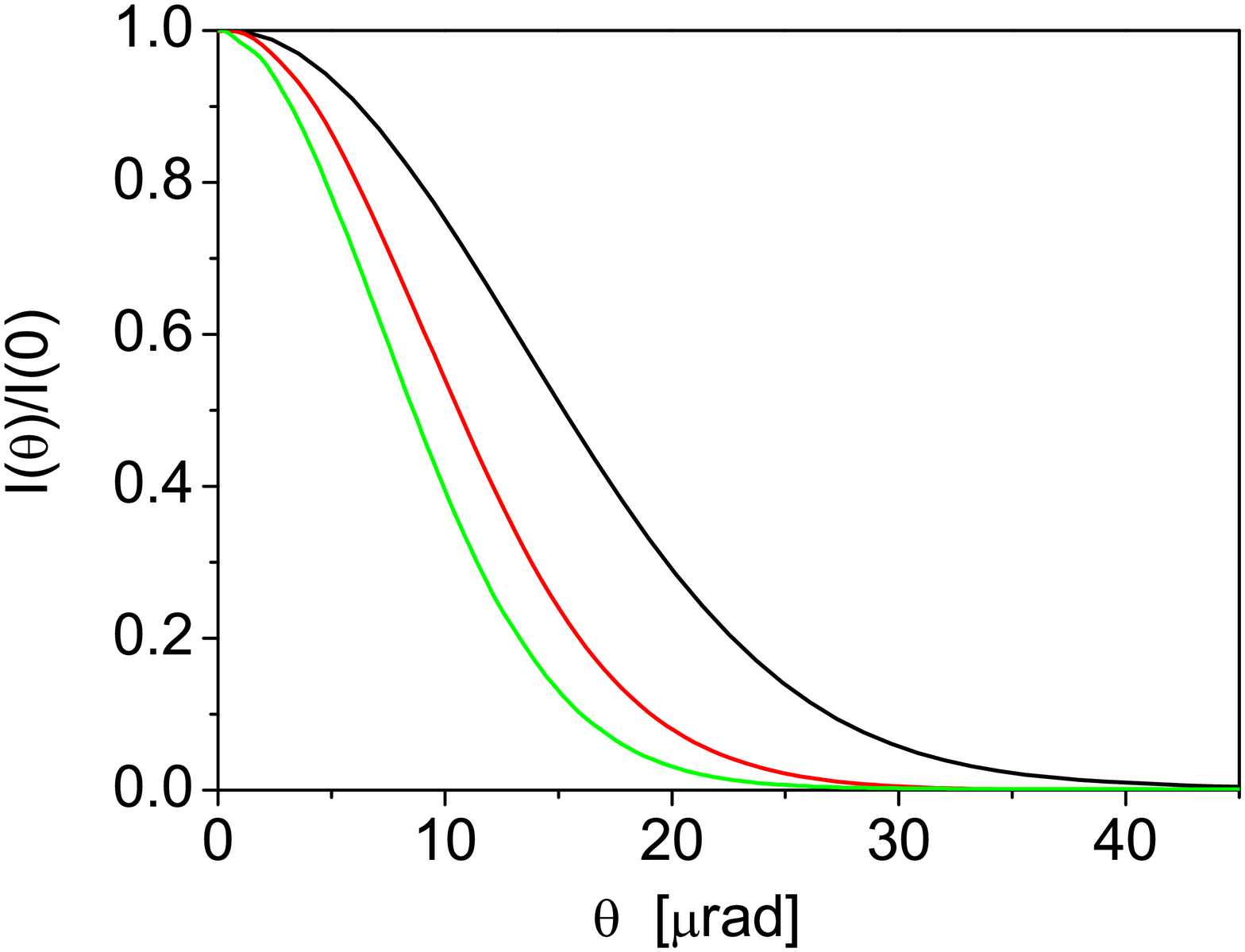}

\caption{
Intensity distribution of the radiation in the saturation point in the near
zone (left plot) and far zone (right plot).
Black, red, and green curves
refer to the 1st, 3rd, and the 5th harmonic, respectively.
Radiation wavelength is 8 nm.
Beta function is 10 m.
Beam current is 1.5 kA.
rms normalized emittance is 1 mm-mrad.
}

\label{fig:nzfz150sat}

\end{figure}

Mode degeneration has significant impact on the pointing stability of SASE FEL.
Let us illustrate this effect with specific example for FLASH operating with
average energy in the radiation pulse of 60 $\mu $J. Left plot in
Fig.~\ref{fig:pz-60uj} shows evolution along the undulator of the radiation
energy in azimuthally symmetric modes and of the energy in the modes with
azimuthal index $n = \pm 1$. Right plot in this figure shows relative
contribution to the total radiation energy of the modes with azimuthal index $n
= \pm 1$. Four consecutive shots are shown here. Temporal profiles of the
radiation pulses are presented in Fig.~\ref{fig:temporal-60uj}. Intensity
distributions in the far zone for these four shots are shown in
four rows in Fig.~\ref{fig:profiles-60uj}. Four profiles on the left-hand side
of each row show intensity distributions in the single slices for the time 40 fs,
50 fs, 60 fs, and 70 fs. Right column presents intensity profiles averaged over
full shots. We see that transverse intensity patterns in slices have rather
complicated shape due to interference of the fields of statistically
independent modes with different azimuthal indexes. Shape of the intensity
distributions changes on a scale of coherence length. Averaging of slice
distributions over radiation pulse results in more smooth distribution.
However, it is clearly seen that
the spot shape of a short radiation pulse changes from pulse to
pulse. The
center of gravity of the radiation pulse
visibly jumps from shot to shot.
Position of the pulse also jumps from shot to shot which is
frequently referred as bad pointing stability.
Note that the effect illustrated here is a fundamental one which takes
place due to the mode degeneration when contribution of the higher azimuthal
modes to the total power is pronouncing (10 to 15\% in our case) . Only in the
case of a long radiation pulse, or after averaging over many pulses we come
asymptotically to an azimuthally symmetric radiation distribution (see
Fig.~\ref{fig:nzfz150sat}). Note that the intensity distributions for the
fundamental harmonic are always wider than those for the higher frequency
harmonics.

\begin{figure}[tb]

\includegraphics[width=0.5\textwidth]{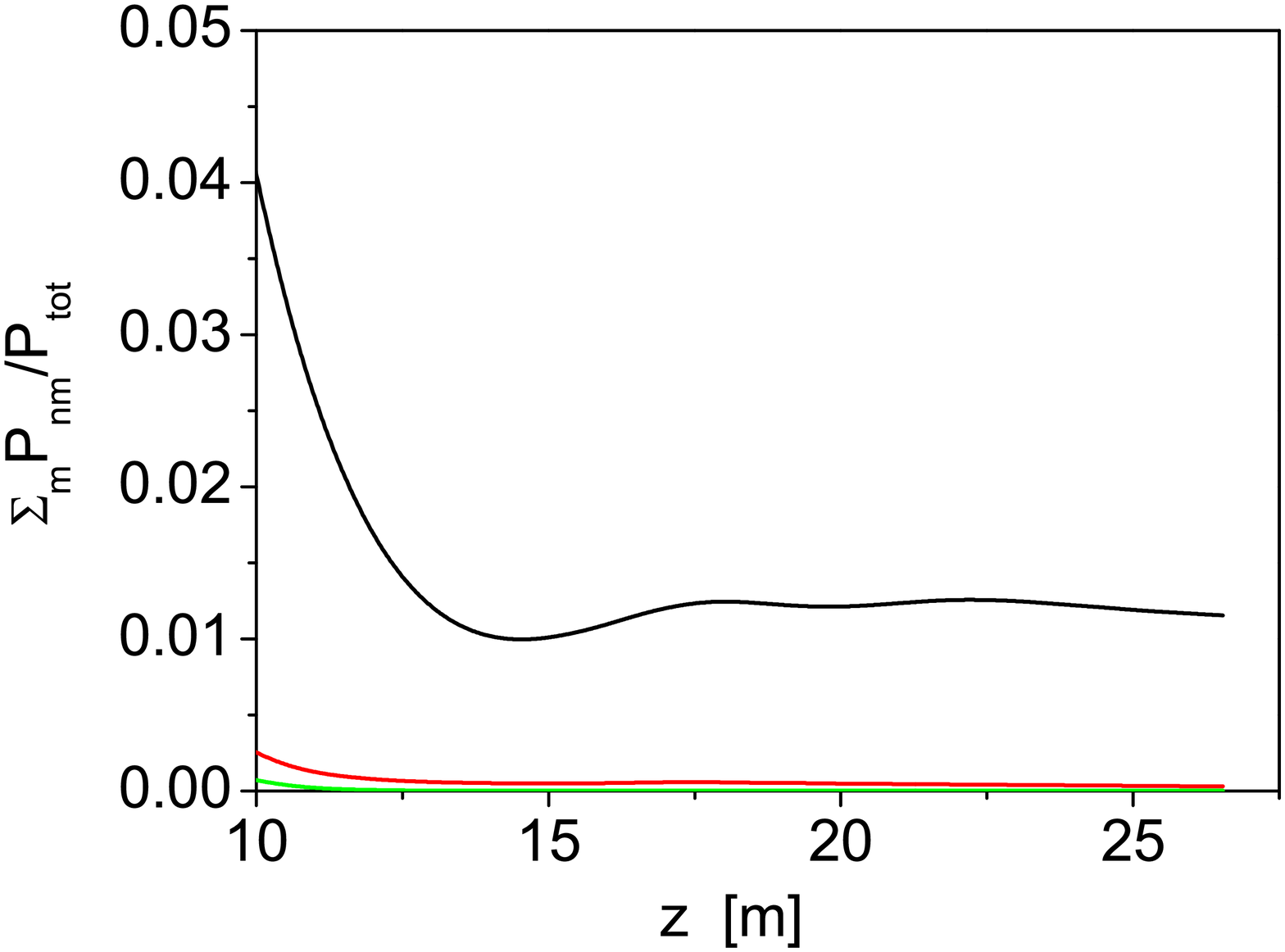}

\caption{
Partial contribution of the higher azimuthal modes of the fundamental harmonic
to the total radiation power.
Black, red, and green
curves refer to the modes with $n = \pm 1$, $n = \pm 2$, and $n = \pm 3$,
respectively.
Radiation wavelength is 8 nm.
Beta function is 5 m.
Beam current is 1 kA.
rms normalized emittance is 0.5 mm-mrad.
}

\label{fig:paz1toptot-bf5i1}

\end{figure}

\begin{figure}[tb]

\includegraphics[width=0.5\textwidth]{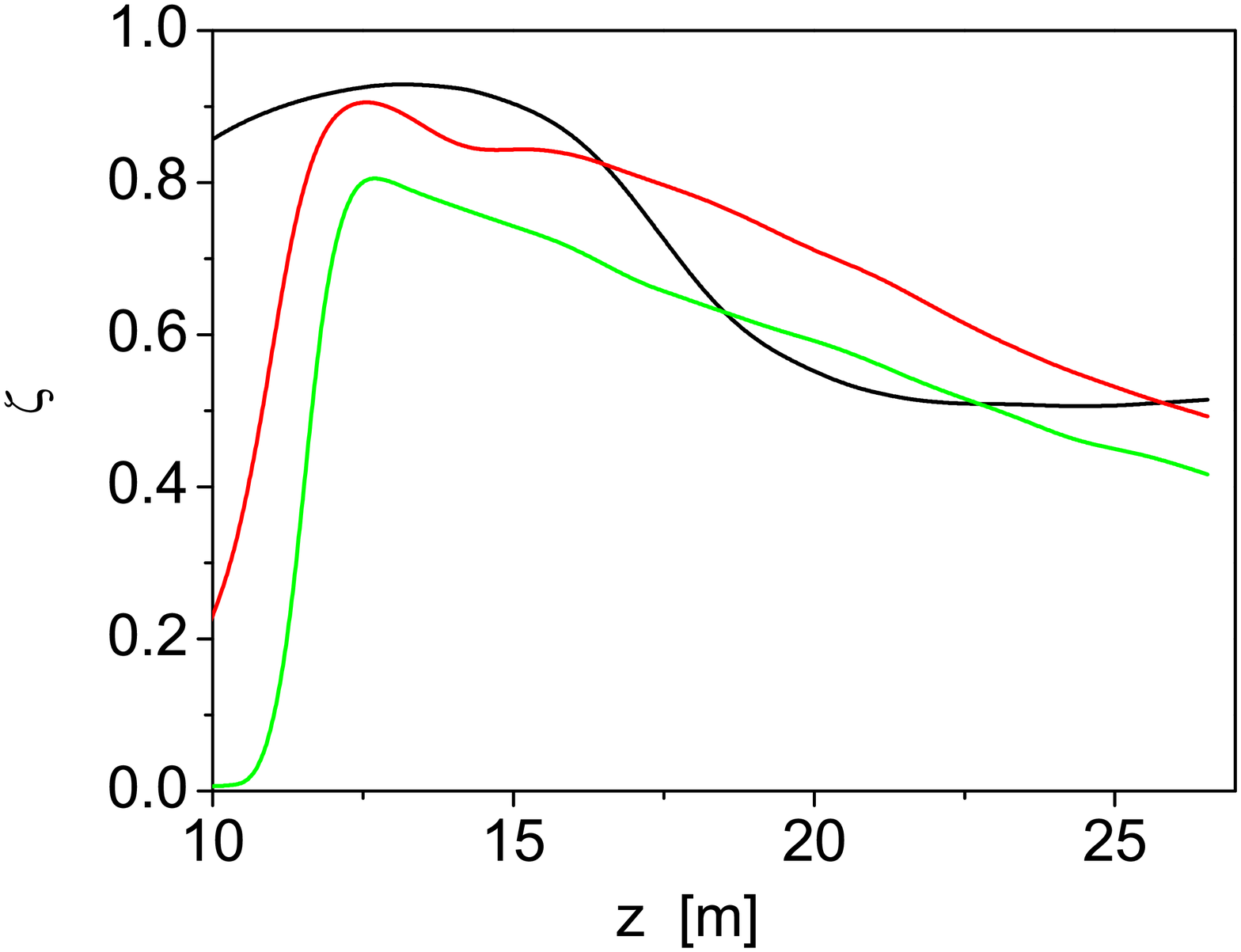}

\caption{Evolution along undulator of the degree of transverse coherence of
the radiation.
Color codes (black, red and green) refer to the 1st harmonic (8nm), the 3rd
harmonic (2.66 nm), and the 5th harmonic (1.6 nm), respectively.
Radiation wavelength is 8 nm.
Beta function is 5 m.
Beam current is 1 kA.
rms normalized emittance is 0.5 mm-mrad.
}
\label{fig:dczbf5i1}
\end{figure}

\section{Discussion}

Simulations presented in this paper trace nearly full range of the parameter
space of FLASH in terms of the emittance and peak current. Detailed
illustration is presented for specific wavelength of 8 nm. We found that there is the fundamental limitation on
pointing stability. Also, the degree of transverse coherence at the saturation point
is visibly smaller than an ultimate value.
This happens because FLASH
FEL operates in the range of physical parameters where different radiation
modes have close values of the gain, i.e. in the parameter range of mode
degeneration. Figure of merit here is the diffraction parameter presenting the
ratio of the electron beam size to the diffraction expansion of the radiation
on a scale of the field gain length \cite{book}. Power of the effect becomes
stronger at the increase of the electron beam size. In the parameter space of
FLASH diffraction parameter is in the range between 10 and 30 (see
Fig.~\ref{fig:bemlamb10i15}). We have shown that for these values of the
diffraction parameter the gain of the first azimuthal mode $TEM_{01}$
approaches to the gain of the ground $TEM_{00}$ mode.
The plot in Fig.~\ref{fig:n1n0b} traces the ratio of the field gain of the first
azimuthal mode $TEM_{01}$ to the gain of the ground FEL mode $TEM_{00}$ versus
radiation wavelength and emittance. We see that situation with mode selection
is unfavorable in the whole wavelength range of FLASH. Ratio is nearly constant
which means that detailed results for 8 nm wavelength can be generalized for
the whole parameter space.

The problem of limited transverse coherence has been discussed already at an early
stage of the project \cite{TESLA-FEL-1995-02} which stimulated the design of
a strong focusing lattice with superimposed focusing in the undulator
\cite{ttf-cdr,pflueger-und-ttf}. This version of the undulator has been used in
the first stage of the project.
Present focusing lattice uses quad doublets installed in the
undulator intersections, and is capable to provide average beta function down
to 5 m \cite{faatz-5m,pflueger-5m}. Currently FLASH operates with the
average focusing function of 10 meters using one quad in the intersection
because this allows to improve the reproducibility of the undulator orbit, thus improving
dramatically FLASH operability.
Otherwise it would be natural to reduce the electron beam size by using the full potential of
the focusing lattice with minimum beta function of 5 m. In addition,
it would desirable to operate FEL at smaller peak current and lower emittance, say
1 kA and 0.5 mm-mrad. In this case the value of the diffraction parameter goes
down to $B = 3.5$ for 8 nm wavelength. This is visibly closer to the
diffraction limit. In addition, mode selection is improved due to higher value
of the betatron motion parameter $\hat{k}_{\beta } = 0.13$. With contour plot
in Fig.~\ref{fig:n1n0b} we can observe the whole range of the parameter space at
FLASH for low current, low beta function mode of operation. Numerical
simulations confirm the results of the mode analysis.
Figure\ref{fig:paz1toptot-bf5i1} shows the contribution of the higher azimuthal
modes to the total power for specific example of emittance 0.5 mm-mrad and peak
current 1 kA . Contribution of the first azimuthal modes falls down to the
value of about 1 \%. Figure~\ref{fig:dczbf5i1} demonstrates that for the
fundamental harmonic the maximal degree of the transverse coherence exceeds
90\%. The degree of transverse coherence is also high for the 3rd and the 5th
harmonic. Note, however, that coherence properties of the radiation degrade
significantly when amplification process enters deep nonlinear regime. The
physics of this phenomena has been discussed in early papers
\cite{coherence-oc,coherence-njp}. Thus, the hint for generating the radiation
with the best coherence properties is to tune the machine such that the
saturation occurs just in the end of the undulator.

We can also use another mechanism for the suppression of the effect
of the mode degeneration. We discussed above that the increase of the energy
spread in the electron beam leads to stronger suppression of higher spatial
modes. Increase of the energy spread can be done with the
laser heater \cite{we-heater}. Features of this effect are demonstrated with
Fig.~\ref{fig:lt2b10}. Increase of the rms energy spread to the value of 0.8
MeV in terms of the mode separation is equivalent to the reduction of the beta
function from 10 to 5 meters. However, the price for this improvement is
significant reduction of the gain of the fundamental mode and of the FEL power,
while reduction of the beta function improves these important FEL parameters.

In view of results obtained we conclude that FLASH (and FLASH2) should be
operated with as strong focusing of the electron beam as technically possible to provide good spatial
coherence and pointing stability of the radiation. Future developments (like design of a new undulator
for FLASH) should also take into account these problems and provide relevant
technical solutions for keeping small size of the electron beam in the
undulator.

\section{Acknowledgement}

We thank Reinhard Brinkmann for many useful discussions.

\end{document}